\documentclass[journal]{IEEEtran}
\usepackage{amsmath,bm,mathtools,commath}
\usepackage[scr=rsfs]{mathalfa}
\usepackage{float}
\usepackage{amssymb}  
\usepackage{graphicx}
\usepackage{epsfig}
\usepackage{epstopdf}
\usepackage{verbatim}
\usepackage{cuted}
\usepackage{multirow,siunitx}
\usepackage{hhline}
\usepackage{url}
\usepackage{cite}
\allowdisplaybreaks
\usepackage{color}
\usepackage{adjustbox,blkarray,bigstrut}
\usepackage{booktabs,array,amsmath,xfrac}
\usepackage{makeidx}         
\usepackage[bottom]{footmisc} 
\usepackage{paralist}
\usepackage{caption,cuted}
\usepackage{subcaption}
\usepackage{dblfloatfix}    
\usepackage{amsthm}

\theoremstyle{plain}
\newtheorem{assumption}{Assumption}[]
\theoremstyle{plain}
\newtheorem{remark}{Remark}[]
\theoremstyle{plain}

\allowdisplaybreaks

\begin{document}
	\title{Model Reduction for Inverters with Current Limiting and Dispatchable Virtual Oscillator Control}
	
	\author{Olaoluwapo~Ajala,~\IEEEmembership{Member,~IEEE,}
		Minghui~Lu,~\IEEEmembership{Member,~IEEE,}
		Brian~Johnson,~\IEEEmembership{Member,~IEEE,}
		Sairaj~Dhople,~\IEEEmembership{Member,~IEEE,}
		and~Alejandro~Dom\'{i}nguez-Garc\'{i}a,~\IEEEmembership{Senior Member,~IEEE}
	}
	
	\maketitle
	
	\begin{abstract}
		This paper outlines reduced-order models for grid-forming virtual-oscillator-controlled inverters with nested current- and voltage-control loops, and current-limiting action for over-current protection. While a variety of model-reduction methods have been proposed to tame complexity in inverter models, previous efforts have not included the impact of current-reference limiting. In addition to acknowledging the current-limiting action, the reduced-order models we outline are tailored to networks with resistive and inductive interconnecting lines. Our analytical approach is centered on a smooth function approximation for the current-reference limiter, participation factor analysis to identify slow- and fast-varying states, and singular perturbation to systematically eliminate the fast states. Computational benefits and accuracy of the reduced-order models are benchmarked via numerical simulations that compare them to higher-order averaged and switched models. 
	\end{abstract}
	
	
	\section{Introduction}	
	
	\IEEEPARstart{T}{he} {increasing deployment of inverter-based resources has altered the dynamic characteristics of electric grids. In this context, there has been significant attention, in recent years, on replacing synchronous power generators with grid-forming (GFM) inverter-based counterparts which, in the absence of synchronous generators, can sustain system voltages and frequency~\cite{Taylor-2016-Fuel,Milano-2018}. Examples of such GFM control strategies include droop~\cite{Chandorkar-1993,Pogaku2007,Zhong_Robust13}, virtual synchronous machines~\cite{Driesen-2008,Zhong2011,Shintai2014,Darco_2015,Liu2017}, and virtual oscillator control~(VOC)~\cite{Gordillo_2002, Johnson2014, Torres2015, Vasquez_2020,Raisz2018, MinghuiSairajBrian2019, Yu2021,Gross2018,TayyebiDorfler2020,Awal2020_unified,Fletcher_ecce,Tobias_2021,Gross_2019,Colombino2019,Lu_IECON2020}.}
	
\begin{figure*}[t!]
		\centering
		\includegraphics{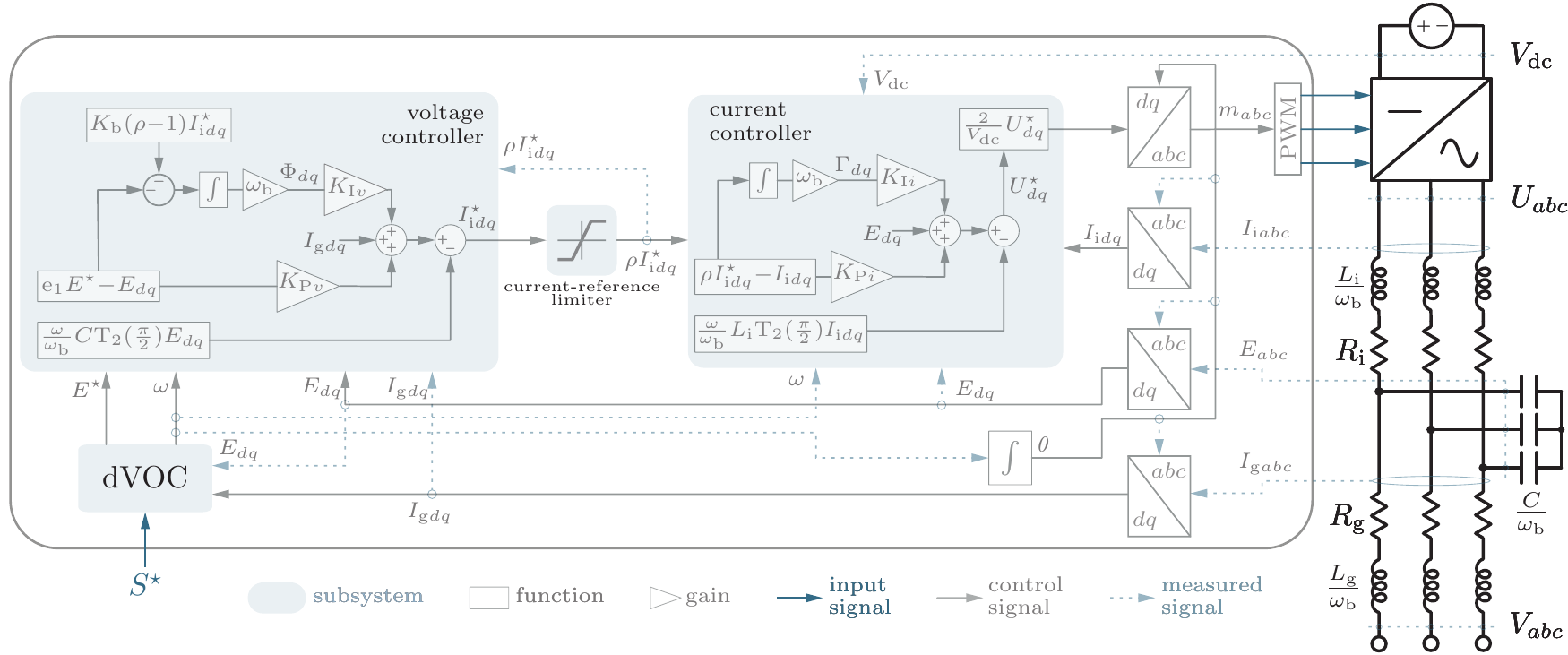}
		\caption{Schematic representation of the GFM inverter per-unit dynamical model with dVOC.}
		\label{fig:AHOinverter}
		\vspace{-0.2in}
	\end{figure*}

	{Motivated by the fact that the VOC strategy is a globally stabilizing control strategy that is able to deal with higher-order harmonics \cite{Sinha2015}}, this paper leverages the theory of singular perturbation~\cite{kokotovicSingular} to outline reduced-order models for a recently proposed variant of VOC called dispatchable virtual oscillator control~(dVOC). The controller leverages the nonlinear dynamics of the Andronov-Hopf oscillator~(AHO) to facilitate synchronization in low-inertia settings, and it also features functionality to respond to power and voltage setpoints~\cite{Fletcher_ecce,Yu2021,Gross2018,MinghuiSairajBrian2019,TayyebiDorfler2020,Awal2020_unified,Tobias_2021,Gross_2019,Colombino2019,Lu_IECON2020}. A schematic representation of the three-phase GFM inverter we examine is sketched in Fig.~\ref{fig:AHOinverter}. The constituent subsystems in the model include:
	\begin{inparaenum}[i)]
		\item the dVOC module that generates voltage and frequency setpoints;
		\item an $LCL$ filter;
		\item a proportional-integral current controller with a  current-reference limiter; and
		\item a proportional-integral voltage controller that includes integrator anti-windup control.\footnote{The structure depicted in Fig.~\ref{fig:AHOinverter} (nested voltage- and current-control loops and current-reference limiting) applies universally to GFM inverters. Distinguishing attributes are introduced by the outermost controller that determines voltage and frequency setpoints. In this work, we assume the outermost controller is implemented via dVOC.}
	\end{inparaenum}
	Two challenges immediately surface when considering the prospect of leveraging such models for system-level analysis. First, such models have multiple dynamic states ($12$ in this particular instance) which presents significant challenges to modeling dynamics of large networks of such inverters with limited computational burden. Second, the nonlinear elements sprinkled throughout the models (stemming from reference-frame transformations, current-reference limits, nonlinear GFM control strategies) present a non-trivial analytical impediment. Notably, these challenges hold true for a broad class of GFM inverter control methods going beyond the dVOC implementation we focus on. 
	
	Our main contribution is the development of reduced-order models for the dVOC flavor of GFM inverters that systematically acknowledge all pertinent nonlinearities in the model, \emph{particularly, the impact of the current-reference limiter}. The current-reference limiter is a key element in the overall control scheme for GFM inverters since it addresses over-currents that would appear otherwise during faults and voltage sags~\cite{Bottrell2014,Sadeghkhani2017,Qoria_2020_current}. It is typically realized with saturation functions that are incompatible with analytical approaches for model reduction. We circumvent this challenge with a smooth-function approximation that carries through the analytical developments. While reduced-order models have been proposed for GFM inverters (we review prior art shortly), to the best of our knowledge, these do not acknowledge current-reference limiters. 
	
	A majority of related literature in model reduction for GFM inverters is centered on droop control~\cite{sairaj2014,Vorobev2018,ajala_springer_2017,Kimball-2015,Caduff_2021}. This is understandable since droop control is one of the earliest proposed GFM control strategies. There are recent efforts---albeit significantly fewer---focused on model reduction for other GFM controls, including virtual synchronous machines~\cite{Dinavahi_2020} and VOC~\cite{Dhople_2018}. These prior efforts have not considered the impact of the current-reference limiter in deriving reduced-order models. This singular aspect underscores the main contribution of our effort. Furthermore, we utilize participation-factor analysis to tease out distinct reduced-order models for dominantly resistive and dominantly inductive interconnecting lines. Such a systematic classification is particularly relevant, since it is well recognized that line attributes have non-trivial impact on system dynamics in low-inertia settings~\cite{Gross2018}. 
	
	{The remainder of this paper is organized as follows. Preliminaries are discussed in Section~\ref{sec:prelim}, the full-order model is overviewed in Section~\ref{sec:full-order}, and reduced-order models are derived in Section~\ref{sec:modelreduct}. Numerical results comparing the reduced-order models with full-order averaged and switched models are presented in Section~\ref{sec:simRes}, and concluding remarks are given in Section~\ref{sec:conclusion}.}
	
	\section{Preliminaries}\label{sec:prelim}
	\subsection{Reference-frame Transformations \& Notation}
	\label{sec:dq}
	Consider the three-phase signal ${f}_{abc}=[f_{a} \, , f_{b} \, , f_{c}]^\top$, where $f_{a}$, $f_{b}$, and $f_{c}$ form a balanced three-phase set. Let $\omega_{\mathrm{b}}$ and $\omega(t)$ denote the nominal angular frequency and the GFM inverter's angular frequency, in \si{\radian\per\second}, respectively, and define
	\begin{equation}
		\theta(t)= \int_0^t\omega(\tau)\,d\tau + \theta(0), \quad
		\delta(t)= \theta(t) - \omega_{\mathrm{b}}t. \label{eqn:angles}
	\end{equation}
	Let ${f}_{{DQ}}=[f_{D} \, , f_{Q}]^\top$ and ${f}_{{dq}}=[f_{d} \, , f_{q}]^\top$ denote transformations of ${f}_{abc}$ to reference frames rotating at angular frequencies $\omega_{\mathrm{b}}$ and $\omega(t)$, respectively. We define the $DQ$ and $dq$ transformations of ${f}_{abc}$ as follows:
	\begin{equation}
		{f}_{DQ} = \mathrm{T}_{1}(\omega_{\mathrm{b}}t){f}_{abc}, \quad {f}_{dq} = \mathrm{T}_{1}(\theta){f}_{abc},\\
	\end{equation}
	where transformation matrix $\mathrm{T}_{1}(\cdot)$ is defined as:
	\begin{equation}
		\begin{split}
			\hspace{-0.025in}\mathrm{T}_{1}(\alpha) =&\ \frac{2}{3}\begin{bmatrix}\cos \alpha\!\! & \cos(\alpha-\frac{2\pi}{3})\!\! & \cos(\alpha+\frac{2\pi}{3})\\ -\sin \alpha \!\! &	-\sin(\alpha-\frac{2\pi}{3}) \!\! & -\sin(\alpha+\frac{2\pi}{3}) \end{bmatrix}.
		\end{split}
	\end{equation}
	Signals in the $DQ$ and $dq$ reference frames are related via
	\begin{equation}
		{f}_{dq} = \mathrm{T}_{2}(\delta){f}_{DQ},    
	\end{equation}
	where rotation matrix $\mathrm{T}_{2}(\cdot)$ is defined as
	\begin{equation}
		\mathrm{T}_{2}(\alpha)= \begin{bmatrix}
			\cos\alpha&\sin\alpha\\
			-\sin\alpha&\cos\alpha
		\end{bmatrix}.
	\end{equation}
	The $2\times 2$ identity matrix is denoted by $\mathrm I$; $\mathrm{e}_1 = [1\,,0]^\top$ and $\mathrm{e}_2 = [0\,,1]^\top$ are the standard basis vectors.
	
	\subsection{Participation Factors}
	For the linear system
	\begin{math}
		\dot{x} = Ax,
	\end{math}
	where $A\in\mathbb{R}^{n\times n}, x\in\mathbb{R}^{n}$, let $r_{ij}$ and $l_{ij}$ denote the $i$-th entries of the right and left eigenvectors associated with the $j$-th eigenvalue of $A$, respectively. The participation of the $i$-th element of $x$ in the $j$-th eigenvalue of $A$ is quantified by
	\begin{equation}
		\frac{|r_{ij}||l_{ij}|}{\sum_{i=1}^n|r_{ij}||l_{ij}|}, \label{eqn:pf1}
	\end{equation} 
	and is called the \emph{participation factor}~\cite{Perez-arriaga1982}. Hereafter, participation factors are normalized using their maximum values. 
	
	\begin{table}[t]
		\centering
		\caption{Model Parameters and Base Values.}
		\resizebox{\columnwidth}{!}{\begin{tabular}{c l c c c r}
				\toprule
				\midrule
				\multirow{1}{*}{Symbol} & Description  & \multicolumn{1}{c}{Value} & Unit & Base value & Base unit\\
				\cmidrule(r){1-2}\cmidrule(l){3-4}\cmidrule(l){5-6}
				{$\psi$} & rotation angle & $\frac{\pi}{4}$ & \si{\radian} & \multirow{3}{*}{N/A} & \multirow{3}{*}{N/A}\\
				\cmidrule(r){1-2}\cmidrule(l){3-4}
				\multirow{2}{*}{$\varepsilon$} & saturation-function & \multirow{2}{*}{\tablenum{0.1}} & \multirow{2}{*}{N/A} & &\\
				& parameter &  &  &  &\\
				\cmidrule(r){1-2}\cmidrule(l){3-4}\cmidrule(l){5-6}
				\multirow{2}{*}{$E_{\mathrm{b}}$} & nominal voltage & \multirow{2}{*}{\tablenum{1}} & \multirow{2}{*}{pu} & \multirow{2}{*}{$\frac{E_r\sqrt{2}}{\sqrt{3}}$} & \multirow{2}{*}{\si{\volt}} \\
				& magnitude (peak)& & & & \\
				\cmidrule(r){1-2}\cmidrule(l){3-4}\cmidrule(l){5-6}
				\multirow{2}{*}{$I_\mathrm{max}$} & peak current & \multirow{2}{*}{\tablenum{1.2}} & \multirow{2}{*}{pu} & \multirow{2}{*}{$\frac{S_r\sqrt{2}}{E_r\sqrt{3}}$} & \multirow{2}{*}{\si{\ampere}} \\
				& limit & & & & \\
				\cmidrule(r){1-2}\cmidrule(l){3-4}\cmidrule(l){5-6}
				\multirow{2}{*}{${L}_{\mathrm{i}}$} & inverter-side & \multirow{2}{*}{\tablenum{0.0196}} & \multirow{2}{*}{pu} & \multirow{7}{*}{$\frac{E_r^2}{S_r\omega_{\mathrm{b}}}$} & \multirow{7}{*}{\si{\henry}} \\
				& inductance &  &  &  &  \\
				\cmidrule(r){1-2}\cmidrule(l){3-4}
				\multirow{4}{*}{${L}_{\mathrm{g}}$} & grid-side & \multirow{2}{*}{\tablenum{0.0196}} & \multirow{2}{*}{pu} & & \\
				& inductance & & & & \\
				\cmidrule(r){2-2}\cmidrule(l){3-4}
				& {grid-side~+~line} & \multirow{2}{*}{\tablenum{0.037}} & \multirow{2}{*}{pu} & & \\
				& inductance & & & & \\
				\cmidrule(r){1-2}\cmidrule(l){3-4}\cmidrule(l){5-6}
				\multirow{2}{*}{$C$} & filter & \multirow{2}{*}{\tablenum{0.1086}} & \multirow{2}{*}{pu} & \multirow{2}{*}{$\frac{S_r}{E_r^2\omega_{\mathrm{b}}}$} & \multirow{2}{*}{\si{\farad}}  \\
				& capacitance & & & & \\
				\cmidrule(r){1-2}\cmidrule(l){3-4}\cmidrule(l){5-6}
				\multirow{2}{*}{${R}_{\mathrm{i}}$} & inverter-side & \multirow{2}{*}{\tablenum{0.0139}} & \multirow{2}{*}{pu} & \multirow{12}{*}{$\frac{E_r^2}{S_r}$} & \multirow{12}{*}{\si{\ohm}} \\
				& resistance & & & & \\
				\cmidrule(r){1-2}\cmidrule(l){3-4}
				\multirow{4}{*}{${R}_{\mathrm{g}}$} & grid-side & \multirow{2}{*}{\tablenum{0.0139}} & \multirow{2}{*}{pu} & & \\
				& resistance & & & & \\
				\cmidrule(r){2-2}\cmidrule(l){3-4}
				& grid-side~+~line & \multirow{2}{*}{\tablenum{0.0313}} & \multirow{2}{*}{pu} & & \\
				& resistance & & & & \\
				\cmidrule(r){1-2}\cmidrule(l){3-4}
				\multirow{2}{*}{$K_{\mathrm{b}}$} & anti-windup & \multirow{2}{*}{\tablenum{0.0347}} & \multirow{2}{*}{pu} & & \\
				& gain & & & & \\
				\cmidrule(r){1-2}\cmidrule(l){3-4}
				\multirow{2}{*}{$K_{\mathrm{P}i}$} & proportional & \multirow{2}{*}{\tablenum{0.9817}} &  \multirow{2}{*}{pu} & & \\  
				& gain & & & & \\
				\cmidrule(r){1-2}\cmidrule(l){3-4}\cmidrule(l){5-6}
				{$K_{\mathrm{I}i}$} & integral gain & \tablenum{0.6944} &  pu & \multirow{3}{*}{$\frac{E_r^2\omega_{\mathrm{b}}}{S_r}$} & \multirow{3}{*}{\si{\per\farad}}\\  
				\cmidrule(r){1-2}\cmidrule(l){3-4}
				\multirow{2}{*}{$\kappa_1$} & synchronization & \multirow{2}{*}{\tablenum{0.0033}} & \multirow{2}{*}{pu} &  & \\
				& gain & & & & \\
				\cmidrule(r){1-2}\cmidrule(l){3-4}\cmidrule(l){5-6}
				\multirow{2}{*}{$K_{\mathrm{P}v}$} & proportional & \multirow{2}{*}{\tablenum{1.4476}} &  \multirow{2}{*}{pu} & \multirow{2}{*}{$\frac{S_r}{E_r^2}$} & \multirow{2}{*}{\si{\per\ohm}}\\
				& gain & & & & \\
				\cmidrule(r){1-2}\cmidrule(l){3-4}\cmidrule(l){5-6}
				{$K_{\mathrm{I}v}$} & integral gain & \tablenum{10.2944} & pu & $\frac{S_r\omega_{\mathrm{b}}}{E_r^2}$ & \si{\per\henry}\\
				\cmidrule(r){1-2}\cmidrule(l){3-4}\cmidrule(l){5-6}
				\multirow{2}{*}{$\omega_\mathrm{bw,i}$} & current-loop & \multirow{2}{*}{\tablenum{50}} & \multirow{2}{*}{pu} & \multirow{4}{*}{$\omega_{\mathrm{b}}$} & \multirow{4}{*}{\si{\radian\per\second}} \\
				& bandwidth & & & & \\
				\cmidrule(r){1-2}\cmidrule(l){3-4}
				\multirow{2}{*}{$\omega_\mathrm{bw,v}$} & voltage-loop & \multirow{2}{*}{\tablenum{13.3333}} & \multirow{2}{*}{pu} & & \\
				& bandwidth & & & & \\
				\cmidrule(r){1-2}\cmidrule(l){3-4}\cmidrule(l){5-6}
				\multirow{2}{*}{$\kappa_2$} & voltage-amplitude & \multirow{2}{*}{\tablenum{0.0796}} & \multirow{2}{*}{pu} & \multirow{2}{*}{$\frac{3\omega_{\mathrm{b}}}{2E_r^2}$} & \multirow{2}{*}{\si{\radian\per\second\per\square\volt}} \\
				& control gain & & & & \\
				\midrule
				\bottomrule
		\end{tabular}}
		\label{tab:param}
	\end{table} 
	
	\section{Averaged Full-order GFM Inverter Model}\label{sec:full-order}
	In this section, we present the averaged full-order dynamical model for GFM inverters covering all control- and physical-layer subsystems.  In subsequent developments, we use per-unit normalization (see, e.g.,~\cite{bergen2000power}) with the rated three-phase power, $S_r$, rated line-to-line voltage (RMS), $E_r$, and nominal system frequency, $\omega_\mathrm b$, serving as base quantities for power, voltage, and frequency, respectively. This facilitates a unified and systematically normalized transcription of all parameters and variables in the dynamic models.
	
	Table~\ref{tab:param} summarizes numerical per-unit values for all the controller and filter parameters adopted in this work, with respective expressions for their base quantities enumerated. Controller and filter parameters are designed assuming $S_r=1500$~\si{\volt}\si{\ampere}, ${E_r}=208$~\si{\volt}, $\omega_{\mathrm{b}}=2\pi60$~\si{\radian\per\second}, and a switching frequency of $10$ k\si{\hertz}. We discuss the design choices of pertinent parameters alongside the overview of each subsystem. 
	
	\subsection{The Dispatchable Virtual Oscillator Controller}\label{subsec:voc}
	Let $P$ and $Q$ denote the active- and reactive-power delivered to the grid at the filter-capacitance terminals, and define $S =[P, Q]^\top$. Furthermore, let $E^{\star}$, ${{P}}^{\star}$, and ${{Q}}^{\star}$ denote references for the voltage magnitude, active power, and reactive power, respectively, and define $S^{\star}=[P^{\star},Q^{\star}]^\top$. Following from the definitions of $\theta$ and $\omega$ in Section~\ref{sec:dq}, the dynamics of frequency and voltage-magnitude references are
	\begin{subequations}
		\begin{align}	
			\dot{\theta} &= \omega = \omega_{\mathrm{b}} + \frac{\omega_{\mathrm{b}} \label{eq:delta} \kappa_1}{(E^{\star})^2} \mathrm{e}_{1}^\top \mathrm{T}_{2}(\psi - \tfrac{\pi}{2})(S^\star - S),
			\\
			\dot{E}^{\star}\! &=\!\! \frac{\omega_{\mathrm{b}}{\kappa_1}}{E^{\star}}  \mathrm{e}_{2}^\top \mathrm{T}_{2}(\psi - \tfrac{\pi}{2})(S^\star - S) \nonumber \\ &\qquad \qquad \qquad + \omega_{\mathrm{b}}\kappa_2(E_{\mathrm{b}}^2-(E^{\star})^2)E^{\star}, \label{eq:Estar}
		\end{align}
	\end{subequations}
	where $E_{\mathrm{b}}$ denotes the nominal inverter voltage magnitude, $\kappa_1$ is the synchronization gain, and $\kappa_2$ is the voltage-amplitude control gain. Furthermore, $\psi\in [0,2\pi)$ denotes the rotation angle of the controller, which is typically tuned based on the $\frac{R}{X}$ ratio of interconnecting lines. For instance, $\psi = \frac{\pi}{2}$ is well suited for inductive transmission lines since it yields active power-frequency and reactive power-voltage droop in steady state~\cite{MinghuiSairajBrian2019}. The model~\eqref{eq:delta}--\eqref{eq:Estar} is built from cycle-averaged dynamics of the Andronov-Hopf oscillator in polar coordinates.\footnote{The dynamical model for the unforced Andronov-Hopf oscillator in polar coordinates takes the general form: $\dot{r} = r (1- r^2)$, $\dot{\theta} = \omega_\mathrm b$. Suitably tailoring this to acknowledge inputs, leveraging periodic-averaging theory, and including pertinent scaling factors, yields~\eqref{eq:delta}--\eqref{eq:Estar}.} 
	
	\subsubsection*{Rationale for choice of $\kappa_1$, $\kappa_2$, and $\psi$} {Numerical values for $\kappa_1$ and $\kappa_2$ listed in Table~\ref{tab:param} ensure that the output voltage and frequency are approximately $0.95$ pu and $59.5$ \si{\hertz}, respectively, when $S=[1,1]^\top$ (full load) and $S^{\star} =[0,0]^\top$. In this context, the output voltage and frequency will return to their nominal values when we set $S^\star =[1,1]^\top$ (see~\cite{MinghuiSairajBrian2019, Lu_IECON2020} for details).} The choice of rotation angle $\psi=\frac{\pi}{4}$ ensures a level of generality by preserving cross-coupling between active power, reactive power, frequency, and voltage.
	
	\subsection{The $LCL$ Filter} \label{subsec:LCL}
	Let ${U}_{dq}$, ${E}_{dq}$, and $V_{DQ}$ denote the filter's inverter-side, capacitor, and grid-side voltages, respectively. Let ${I}_{\mathrm{i}dq}$ and ${I}_{\mathrm{g}dq}$ denote the filter's inverter-side and grid-side currents, respectively. The filter dynamics are captured by 
	\begin{subequations}
		\begin{align}
			\dot{{I}}_{\mathrm{i}dq} &= \Big(\omega \mathrm{T}_{2}(\tfrac{\pi}{2}) - \omega_\mathrm b\frac{R_\mathrm i}{L_\mathrm i}\mathrm I\Big) I_{\mathrm{i}dq} + \frac{\omega_\mathrm b}{L_\mathrm i} (U_{dq} - E_{dq}),\label{eq:LCL_filter1} \\
			\dot{E}_{dq} &= \omega \mathrm{T}_{2}(\tfrac{\pi}{2}) E_{dq} + \frac{\omega_\mathrm b}{C} (I_{\mathrm{i}dq} - I_{\mathrm{g}dq}), \label{eq:LCL_filter2}\\
			\dot{{I}}_{\mathrm{g}dq} &= \Big(\omega \mathrm{T}_{2}(\tfrac{\pi}{2}) - \omega_\mathrm b\frac{R_\mathrm g}{L_\mathrm g}\mathrm I\Big) I_{\mathrm{g}dq} \nonumber\\ &\qquad \qquad + \frac{\omega_\mathrm b}{L_\mathrm g} ( E_{dq} - \mathrm{T}_{2}(\delta)V_{DQ}), \label{eq:LCL_filter3}
		\end{align}
	\end{subequations}
	where ${L}_{\mathrm{i}}$, ${L}_{\mathrm{g}}$, and $C$ denote the inverter-side inductance, grid-side inductance, and capacitance of the $LCL$ filter, respectively; ${R}_{\mathrm{i}}$ and ${R}_{\mathrm{g}}$ denote the non-ideal series resistances associated with the inverter- and grid-side inductors, respectively.
	
	\subsubsection*{Rationale for choice of $L_\mathrm{g}$, $L_\mathrm{i}$, and $C$} {One well-established approach for designing the LCL filter involves selecting a resonant frequency that is between  ten times the value of the grid frequency ($60$~\si{\hertz} in this case) and half the value of the switching frequency ($10$~k\si{\hertz} in this case) \cite{Reznik2014_LCL}}. For a given resonant frequency ($1.8$~k\si{\hertz} in our design), picking the inverter-side inductance to be equal to the grid-side inductance ensures the smallest capacitive reactive power~\cite{Channegowda2010}. These design considerations yield the choice of values for $L_\mathrm{g}$, $L_\mathrm{i}$, and $C$ reported in Table~\ref{tab:param}. The parasitic resistances follow from the hardware prototype realization discussed in~\cite{MinghuiSairajBrian2019}.
	
	\begin{remark} [Defining Active- and Reactive-power Outputs]
		With the filter currents and voltages formally annotated, the active- and reactive-power outputs measured at the filter capacitors can be expressed as:
		\begin{equation}
			\begin{split}
				P = E_{dq}^\top I_{\mathrm{g}dq},\quad	Q = E_{dq}^\top \mathrm{T}_{2}(-\tfrac{\pi}{2}) I_{\mathrm{g}dq}.\label{eq:PQ}
			\end{split}
		\end{equation}	
	\end{remark}
	
	\subsection{The Current-reference Limiter}\label{subsec:i-ctrllimit}
	As depicted in Fig.~\ref{fig:AHOinverter}, the current controller acts on a reference command, denoted by ${I}^{\star}_{\mathrm{i}dq}$, which is generated by the voltage controller (to be described in detail later). As a first step, the magnitude of this reference is saturated to the inverter peak-current limit, $I_\mathrm{max}$. In the literature, this has been accomplished with the following saturation function~\cite{TaulBlaabjerg2020,TayyebiDorfler2020}:
	\begin{equation}
		\min\left(1,\frac{I_\mathrm{max}}{\| I_{\mathrm{i}dq}^\star\|_2}\right){I}^{\star}_{\mathrm{i}dq}. \label{eq:min}
	\end{equation}
	For analytical convenience, we model the current-reference limiting operation via the product $\rho I^\star_{\mathrm{i}dq}$, where $\rho$ is given by:
	\begin{equation}
		\begin{split} \label{eq:rho}
			\rho =-\varepsilon\ln\Bigg(\exp\Big(\frac{-1}{\varepsilon}\Big)+\exp\Big(\frac{-I_\mathrm{max}}{\varepsilon\| I_{\mathrm{i}dq}^\star\|_2}\Big)\Bigg). 
		\end{split}
	\end{equation}
	The $\min(\cdot,\cdot)$ function in~\eqref{eq:min} can be approximated by $\rho$ for small values of the saturation-function parameter, $\varepsilon$.
	
	\subsubsection*{Rationale for choice of $I_\mathrm{max}$ and $\varepsilon$} 
\begin{figure}[t!]
		\centering
		\includegraphics{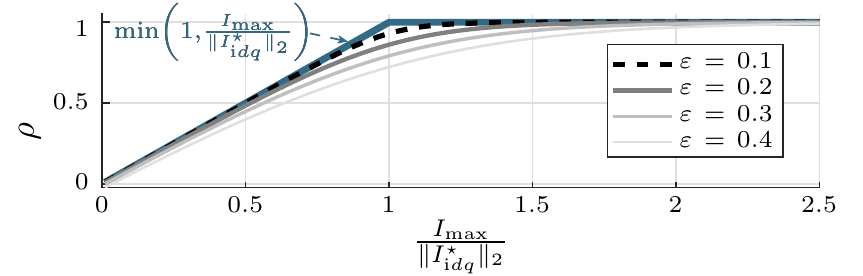}
		\caption{Approximations of the $\min(\cdot,\cdot)$ function~\eqref{eq:min} with $\rho$~\eqref{eq:rho} for $\varepsilon=0.1, 0.2, 0.3$, and  $0.4$.}
		\label{fig:saturation}
		\vspace{-0.2in}
	\end{figure}

The choice of $I_\mathrm{max}$ in Table~\ref{tab:param} follows \cite{TaulBlaabjerg2020}, which has espoused limits in this range for GFM inverters. The choice of $\varepsilon$ ensures a close match of the approximation $\rho$ to the $\mathrm{min}(\cdot,\cdot)$ function. (See Fig.~\ref{fig:saturation}.) 
	\subsection{The Voltage Controller}\label{subsec:v-ctrl}
	The voltage controller acts on the terminal-voltage reference, denoted by $E^\star$, which is generated by the dVOC module and yields the current-control reference command, $I_{{\mathrm i}dq}^\star$. Let $\Phi_{dq}$ denote the state variable of the voltage controller. The dynamics of the voltage controller are governed by
	\begin{subequations}
		\begin{align}
			\dot{\Phi}_{dq}
			&= \omega_{\mathrm{b}}(\mathrm{e}_{1}{E}^{\star}-{E}_{dq})+\omega_{\mathrm{b}}K_{\mathrm{b}}(\rho-1)
			{I}_{\mathrm{i}dq}^{\star},\label{eq:Vcont1}\\
			\hspace{-0.025in}\widehat{\rho} {I}^{\star}_{{\mathrm{i}}dq}
			&= \frac{K_{\mathrm{P}v}}{\omega_\mathrm b} \dot{\Phi}_{dq}+ K_{\mathrm{I}v}
			{\Phi}_{dq} + {I}_{\mathrm{g}dq} - \frac{\omega}{\omega_\mathrm b}C \mathrm{T}_{2}(\tfrac{\pi}{2})E_{dq},\label{eq:Vcont2}
		\end{align}
	\end{subequations}
	where $\widehat{\rho}= 1+ {K_{\mathrm{P}v}K_\mathrm{b} (\rho-1)}$, $K_{\mathrm{P}v}$ and $K_{\mathrm{I}v}$ denote the voltage controller's proportional and integrator gains, respectively, and $K_{\mathrm{b}}$ denotes the integrator anti-windup gain. In effect,~\eqref{eq:Vcont1} and~\eqref{eq:Vcont2} constitute a proportional-integral (PI) control system with reference $E^\star$, controlled signal $E_{dq}$, and control input ${I}^{\star}_{{\mathrm{i}}dq}$. {The loop includes a feed-forward compensation term, ${I}_{\mathrm{g}dq}-\frac{\omega}{\omega_\mathrm b}C \mathrm{T}_{2}(\tfrac{\pi}{2})E_{dq}$, and an integrator anti-windup term $K_{\mathrm{b}}(\rho-1)
		{I}_{\mathrm{i}dq}^{\star}$ \cite{TaulBlaabjerg2020}\cite[pp. 245--262]{yazdaniVSC}.}
	
	\subsubsection*{Rationale for choice of $K_{\mathrm{P}v}$, $K_{\mathrm{I}v}$, and $K_\mathrm b$} The bandwidth of the current-control loop (which we denote by $\omega_\mathrm{bw,i}$) is typically one-tenth to one-fifth of the switching frequency, and that of the voltage-control loop (which we denote by $\omega_\mathrm{bw,v}$) is multiple times slower for time-scale separation. The PI gains for the controller are set as $K_{\mathrm{P}v} = \omega_\mathrm{bw,v} C$ and $K_{\mathrm{I}v} = 2K_{\mathrm{P}v}\omega_\mathrm{bw,v}^2/\omega_\mathrm{bw,i}$ to yield an approximately first-order response~\cite[pp. 253--257]{yazdaniVSC}. A manual trial-and-error process is used to tune the value of $K_\mathrm b$.
	
	\subsection{Current Controller} \label{subsec:i-ctrl}
	Let $\Gamma_{dq}$ denote the state variable corresponding to the current controller, and ${U}^{\star}_{dq}$ denote the output of the current controller. Note that $U^\star_{dq}$ generates the PWM reference signals for the inverter. The  dynamics of the current controller are
	\begin{subequations}
		\begin{align}
			\dot{\Gamma}_{dq} &= \omega_\mathrm b (\rho I^\star_{\mathrm{i}dq} - I_{\mathrm{i}dq}), \label{eq:Gamma}\\
			U^{\star}_{dq} &=  \frac{K_{\mathrm{P}i}}{\omega_\mathrm b}\dot{\Gamma}_{dq} + K_{\mathrm{I}i}
			{\Gamma}_{dq}
			+
			{E}_{dq} - \frac{\omega}{\omega_\mathrm b}L_\mathrm i \mathrm{T}_{2}(\tfrac{\pi}{2})
			{I}_{\mathrm{i}dq}, \label{eq:Ustar}
		\end{align}
	\end{subequations}
	where $K_{\mathrm{P}i}$ and $K_{\mathrm{I}i}$ are the current-controller proportional and integrator gains, respectively. In effect,~\eqref{eq:Gamma}--\eqref{eq:Ustar} close the loop around the inverter-side inductor current, $I_{\mathrm{i}dq}$, with a PI loop that acts on a saturated version of the current reference, $\rho I^\star_{\mathrm{i}dq}$. {The feed-forward compensation term, $E_{dq}- \frac{\omega}{\omega_\mathrm b}L_\mathrm i \mathrm{T}_{2}(\tfrac{\pi}{2})
		{I}_{\mathrm{i}dq}$,
		in~\eqref{eq:Ustar} enhances disturbance rejection~\cite[p.~219]{yazdaniVSC}.}
	
	\subsubsection*{Rationale for choice of $K_{\mathrm{P}i}$ and $K_{\mathrm{I}i}$} The PI gains are set as $K_{\mathrm{P}i} = \omega_\mathrm{bw,i} L_\mathrm{i}$ and $K_{\mathrm{I}i}= \omega_\mathrm{bw,i} R_\mathrm{i}$ to yield an approximately first-order response from reference to output~\cite[p.~247]{yazdaniVSC}.
	
	\subsection{The Three-phase Inverter}\label{subsec:3_invl}
	The three-phase line-neutral voltage at the inverter switch terminals is captured by $U_{abc} = \frac{V_\mathrm{dc}}{2} m_{abc}$, where $m_{abc}$ denotes the pulse-width modulation (PWM) signals and $V_\mathrm{dc}$ is the per-unitized dc-side voltage (see \cite{yazdaniVSC}, pp.~115--126 for details). With the control architecture sketched in Fig.~\ref{fig:AHOinverter}, it emerges that the averaged voltages at the inverter switched terminals in the $dq$ reference frame are given by: $U_{dq}=\ U_{dq}^{\star}$.
	
	\begin{figure*}[t!]
		\begin{subfigure}[b]{\columnwidth}
			\centering
			\includegraphics{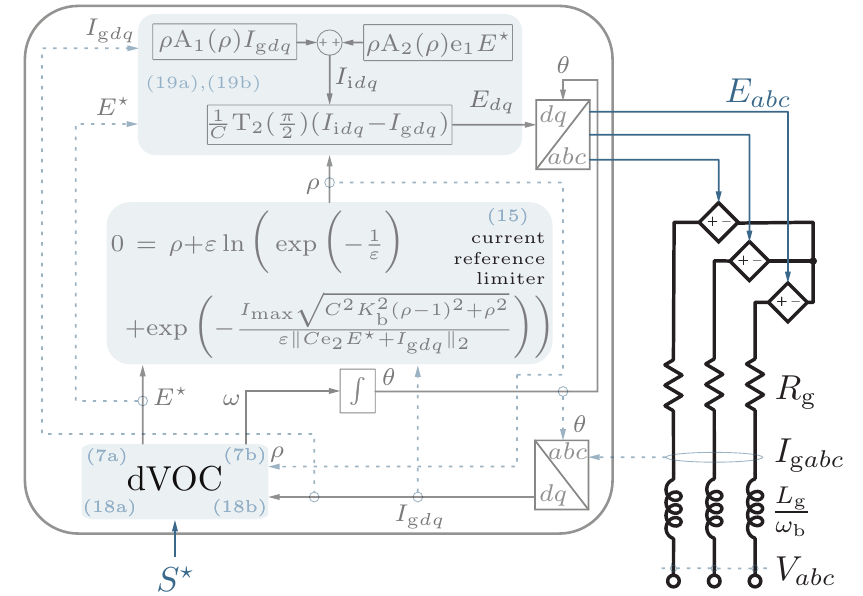}
			\caption{Inductive interconnecting lines.}
			\label{fig:reducedorder_L}
		\end{subfigure}
		\hfill
		\begin{subfigure}[b]{\columnwidth}
			\centering
			\includegraphics{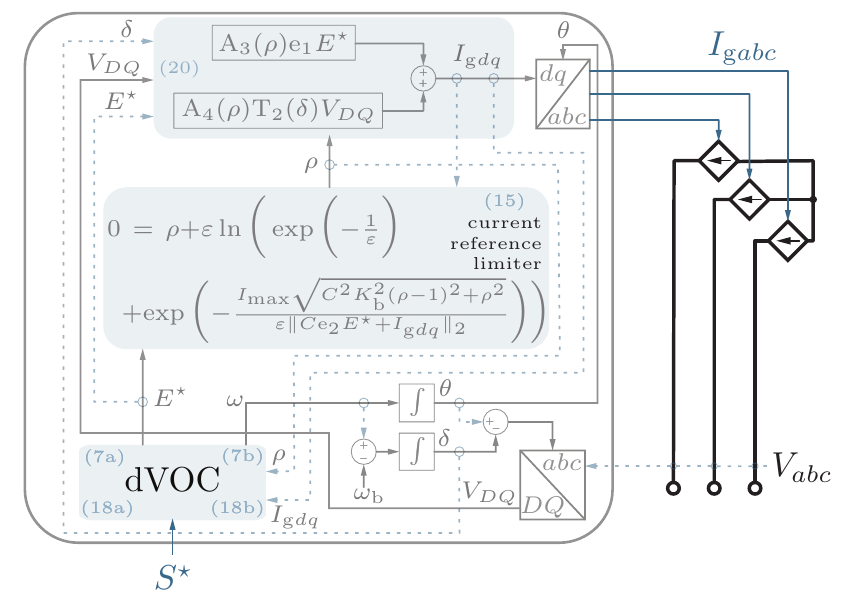}
			\caption{Resistive interconnecting lines.}
			\label{fig:reducedorder_R}
		\end{subfigure}
		\caption{Reduced-order models for GFM inverters with a current-reference limiter and dVOC.}
		\label{fig:reduced_models} \vspace{-0.2in}
	\end{figure*}
	\begin{figure*}[t!]
		\centering
		\includegraphics{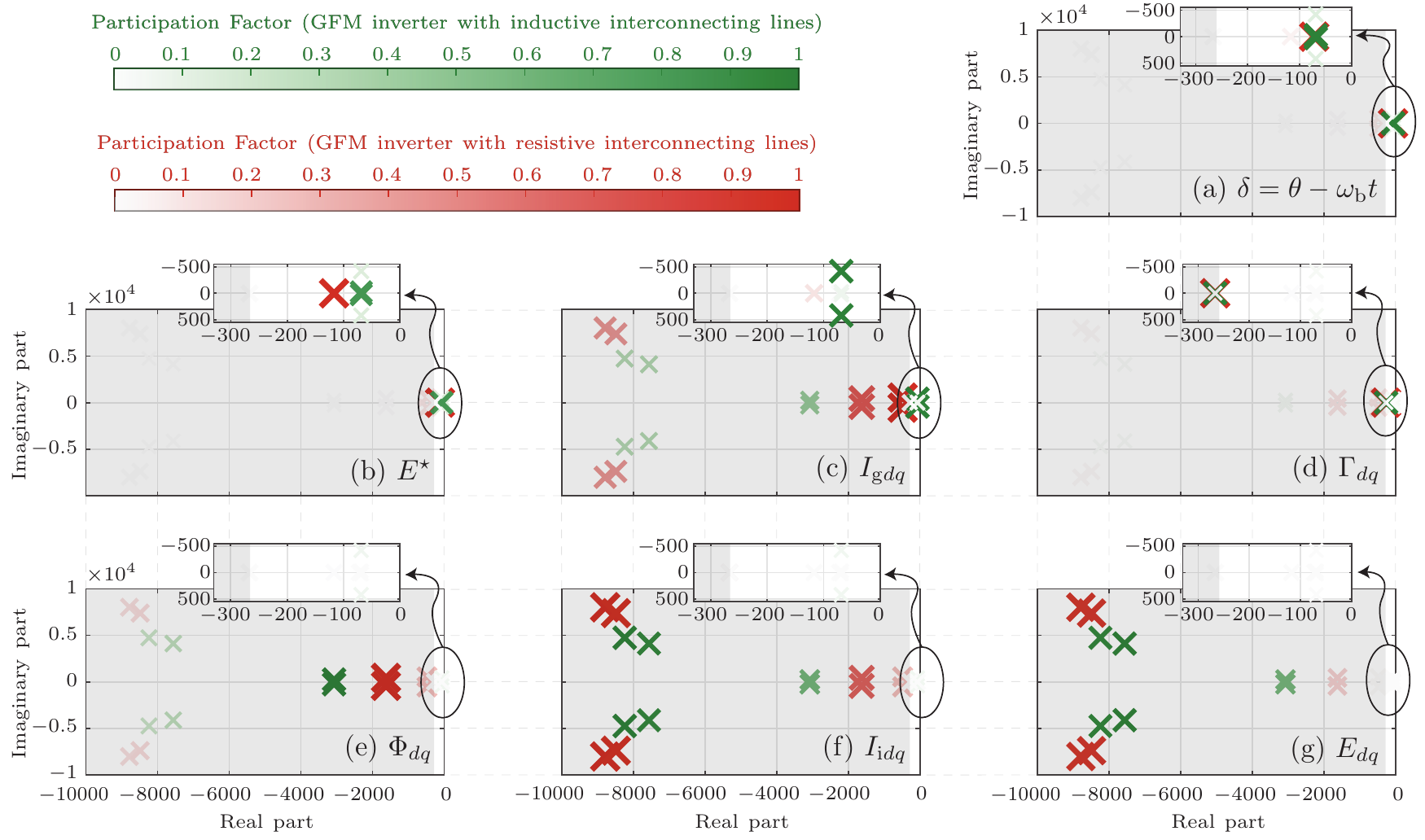}
		\caption{{Eigenvalues of the averaged full-order model's Jacobian matrix, with each eigenvalue color coded according to the participation factor of: (a)~$\delta= \theta - \omega_{\mathrm{b}}t,\;$ (b)~$E^{\star},\;$ (c)~${I}_{\mathrm{g}dq},\;$ (d)~$\Gamma_{dq}\;$ (e)~$\Phi_{dq},\;$ (f)~${I}_{\mathrm{i}dq},\;$ and (g)~${E}_{dq}$. Elements of each $dq$ variable have identical participation factors. The case of inductive (resistive) interconnecting lines is colored green (red). The region shaded in gray identifies eigenvalues associated with fast states.}} 
		\label{fig:EigenPartFac}
	\end{figure*}
	\section{The Per-unit Reduced-order Models}
	\label{sec:modelreduct}
	In this section, we present the main results of this work, namely: reduced-order models for GFM inverters with dVOC acknowledging the current-reference limiter. The reduced-order models are sketched in Figs.~\ref{fig:reducedorder_L} and~\ref{fig:reducedorder_R}, for inverters with inductive and resistive interconnecting lines, respectively. Compared to Fig.~\ref{fig:AHOinverter}, in both cases, the dynamics of the current- and voltage-control loops are abstracted, as are the dynamics corresponding to the inverter-side inductance, $L_\mathrm{i}$, and filter capacitance, $C$. The impact of the current-reference limiter is preserved through an algebraic constraint in both cases. 
	
	We begin with a discussion on how to determine the dimension of the reduced-order models. Subsequently, we discuss the derivation of the reduced-order models and consider the special case without the current-reference limiter. 
	
	\subsection{Determining the Order of the Reduced-order Models} \label{sec:OrderIntro}
	The first step in model-order reduction is to identify \emph{fast} and \emph{slow} states in the inverter dynamical model while acknowledging that the inverter could be interconnected to a network with dominantly inductive \emph{or} resistive lines. To aid this, in Fig.~\ref{fig:EigenPartFac}, we plot eigenvalues and participation factors corresponding to a linearized version of the inverter dynamical model for parameters presented in Table~\ref{tab:param} and inputs: $S^{\star}= [2,2]^\top$~pu, ${V}_{DQ}= [1,0]^\top$~pu. Two sets of results are plotted for inductive and resistive interconnecting lines, and they are distinguished based on the values of $R_\mathrm g$ and $L_\mathrm g$ utilized in the simulation. For the case with inductive (resistive) lines, we set $L_{\mathrm{g}}$ ($R_{\mathrm{g}}$) equal to the sum of the grid-side inductance (resistance) and the line inductance (resistance). 
	
	A careful examination of Fig.~\ref{fig:EigenPartFac} reveals that a  bandwidth of (approximately) $260$~\si{\radian\per\second} is a reasonable cut-off to separate slow and fast states for both inductive and resistive interconnections. The region shaded in gray in Fig.~\ref{fig:EigenPartFac} identifies eigenvalues whose real parts take values less than $-260$~\si{\radian\per\second}, and should be considered as fast dynamics. The choice of the cut-off bandwidth is determined based on the locus of (the real-part of) eigenvalues most influenced by fast states that are not as strongly dependent on the nature of interconnecting lines. A careful examination of Fig.~\ref{fig:EigenPartFac} suggests that in this case, these fast states are $\Gamma_{dq}$, i.e., the states associated with the current controller.\footnote{The real-part of eigenvalues most impacted by $\Gamma_{dq}$, emphasized in Fig.~\ref{fig:EigenPartFac}(d), is $-266.7$~\si{\radian\per\second} for both cases considered. We establish the cut-off frequency to be $260$~\si{\radian\per\second} to go with a well-rounded number.} The inferences reported above hold for a wide range of inputs ($S^\star,V_{DQ}$) and line parameters ($R_\mathrm g, L_\mathrm g$).  
	
	\subsection{Procedure Involved in Model-order Reduction} \label{subsec:ModRed}
	The results presented in this subsection are based on singular perturbation analysis~\cite{kokotovicSingular}. We require the following reasonable assumption for the results to hold:
	\begin{assumption}\label{assum:speed}
		The GFM inverter's angular frequency, ${\omega}$, satisfies the constraint 
		\begin{equation*}
			\Big|\frac{\omega-{\omega}_{\mathrm{b}}}{{\omega}_{\mathrm{b}}}\Big|\leq\epsilon,    
		\end{equation*}
		where $\epsilon$ is a dimensionless parameter.
	\end{assumption}\noindent
	\noindent For a nominal frequency, $\omega_\mathrm b = 60~\si{\hertz}$, and with the choice $\epsilon = 260^{-1}$ (motivated by the discussion on cut-off bandwidth in Section~\ref{sec:OrderIntro}), the above assumption implies that the reduced-order models that follow are valid when the GFM inverter's frequency is within $59.77$~\si{\hertz} and $60.23$~\si{\hertz}.
	
	The following steps, which are described in detail in the Appendix, are involved in model reduction:
	\begin{enumerate}[1.]
		\item The dynamics introduced in Section~\ref{sec:full-order} are represented compactly as a $12$th-order set of differential equations by:
		{\begin{enumerate}[(i)]
				\item substituting (12a) into (12b) and solving for ${I}^{\star}_{{\mathrm{i}}dq}$,
				\item substituting the result into (11), (12a) and (13a)  (equation (12b) is no longer needed),
				\item substituting the new expression for (13a) into (13b),
				\item using the new expression for (13b) to obtain an expression for $U_{dq}$ (recall that $U_{dq}=\ U_{dq}^{\star}$), and substituting this into (8a) (equation (13b) is no longer needed).
		\end{enumerate}}
		\item The right- and left-hand sides of the differential equations for fast states are multiplied by $\epsilon = 260^{-1}$. Resulting equations are in the standard singular perturbation form:
		\begin{subequations}
			\begin{align} \dot{x}=&\ f\left({x},{z},\epsilon\right),\label{eqn:modelx_HO}\\	\epsilon\dot{z}=&\ g\left({x},{z},\epsilon\right),\label{eqn:modelz_HO}
			\end{align}\label{eqn:model_HO}
		\end{subequations} where $f\left(\cdot,\cdot,\cdot\right)$ and $g\left(\cdot,\cdot,\cdot\right)$ are continuously differentiable functions of their arguments, the elements of $x$ are the slow-varying states, and the elements of $z$ are the fast-varying states.
		From the discussion in Section~\ref{sec:OrderIntro}, we note that when inductive lines interconnect the GFM inverter to the bus, ${x} =
		[\theta,E^{\star},{I}_{{\mathrm{g}}dq}^\top ]^\top$ and ${z} = [ {I}_{\mathrm{i}dq}^\top,{E}_{dq}^\top,\Phi_{dq}^\top,\Gamma_{dq}^\top ]^\top$, and \eqref{eqn:model_HO} is recovered by 
		{collecting the new expressions obtained for (8a), (8b), (12a), and (13a) and multiplying the left- and right-hand sides of each equation by $\epsilon$.} On the otherhand, when resistive interconnecting lines are used, ${x} = [\theta, E^{\star}]^\top$ and ${z} = [{I}_{{\mathrm{g}}dq}^\top , {I}_{\mathrm{i}dq}^\top,{E}_{dq}^\top, \Phi_{dq}^\top,\Gamma_{dq}^\top ]^\top$, and \eqref{eqn:model_HO} is recovered by {collecting the new expressions obtained for (8a), (8b), (8c), (12a), and (13a) and multiplying the left- and right-hand sides of each equation by $\epsilon$}.
		\item The set of differential equations comprising~\eqref{eqn:modelz_HO} are replaced with algebraic counterparts. We employ a zero-order approximation of the integral manifold for ${z}$ as the algebraic counterpart; this is derived by setting $\epsilon=0$ on the left-hand side of~\eqref{eqn:modelz_HO}, setting $\frac{\omega-{\omega}_{\mathrm{b}}}{{\omega}_{\mathrm{b}}}=0$ in the resulting set of equations (this follows from Assumption~\ref{assum:speed}), and solving for ${z}$ as a function of ${x}$. The resulting set of equations yield a $4$th- ($2$nd-) order model for GFM inverters with inductive (resistive) lines.
	\end{enumerate}
	
	\subsection{Reduced-order Models}\label{sec:reduced-models}
	Following the steps above, which are described in detail in the Appendix, it emerges that the action of the current-reference limiter is captured by solving for $\rho$ in
	\begin{align}
		\begin{split}
			0 &= {\rho}+\varepsilon\ln\Bigg(\exp\Big(-\frac{1}{\varepsilon}\Big)  \\
			&+\exp\Big(-\frac{I_\mathrm{max}{\sqrt{C^2K_{\mathrm{b}}^2({\rho} -1)^2 +{\rho} ^2}}}{\varepsilon\| C\mathrm{e}_2{E}^{\star}+I_{\mathrm{g}dq}\|_2}\Big)\Bigg).\label{eq:rho4thorder}
		\end{split}
	\end{align}
	Compared to~\eqref{eq:rho}, the above algebraic constraint is recognizably cumbersome; however, it captures the impact of current-reference limiting in an analytically tractable fashion.
	
	We summarize the differential and algebraic equations corresponding to the two sets of reduced-order models next. Before doing so, we will find the below definitions useful: 
	\begin{subequations}
		\begin{align}
			\mathrm{A}_{1}(\rho) &= 
			\begin{bmatrix} \frac{\rho}{C^2K_{\mathrm{b}}^2({\rho} -1)^2 +{\rho} ^2} & - \frac{CK_{\mathrm{b}}({\rho} -1)}{C^2K_{\mathrm{b}}^2({\rho} -1)^2 +{\rho} ^2}\\
				\frac{CK_{\mathrm{b}}({\rho} -1)}{C^2K_{\mathrm{b}}^2({\rho} -1)^2 +{\rho} ^2} & \frac{\rho}{C^2K_{\mathrm{b}}^2({\rho} -1)^2 +{\rho} ^2}
			\end{bmatrix}, \label{eq:Tirho} \\
			\mathrm{A}_{2}(\rho) &= 
			\begin{bmatrix} -\frac{C^2K_{\mathrm{b}}({\rho} -1)}{C^2K_{\mathrm{b}}^2({\rho} -1)^2 +{\rho} ^2} & -\frac{C\rho}{C^2K_{\mathrm{b}}^2({\rho} -1)^2 +{\rho} ^2}\\
				\frac{C\rho}{C^2K_{\mathrm{b}}^2({\rho} -1)^2 +{\rho} ^2} & -\frac{C^2K_{\mathrm{b}}({\rho} -1)}{C^2K_{\mathrm{b}}^2({\rho} -1)^2 +{\rho} ^2}
			\end{bmatrix}, \label{eq:Tvrho} \\
			\mathrm{A}_{3}(\rho) &= 
			\begin{bmatrix} \frac{f_{1}(\rho)}{f_{5}(\rho)} & -\frac{f_{2}(\rho)}{f_{5}(\rho)}\\
				\frac{f_{2}(\rho)}{f_{5}(\rho)} & \frac{f_{1}(\rho)}{f_{5}(\rho)}
			\end{bmatrix}, \quad
			\mathrm{A}_{4}(\rho) = 
			\begin{bmatrix} \frac{f_{3}(\rho)}{f_{5}(\rho)} & - \frac{f_{4}(\rho)}{f_{5}(\rho)}\\
				\frac{f_{4}(\rho)}{f_{5}(\rho)} & \frac{f_{3}(\rho)}{f_{5}(\rho)}
			\end{bmatrix}, \label{eq:Tgvrho}
		\end{align} 
	\end{subequations}
	where, we introduce:
	\begin{align*}
		f_{1}(\rho) &= (CL_\mathrm g-1)K_{\mathrm{b}}{\rho}({\rho} -1)+R_\mathrm g{\rho}^2, \nonumber \\
		f_{2}(\rho) &= CR_\mathrm g K_{\mathrm{b}}{\rho}({\rho} -1)-L_\mathrm g{\rho}^2, \nonumber \\
		f_{3}(\rho) &= K_{\mathrm{b}}\rho(\rho-1)-R_\mathrm g (C^2K_{\mathrm{b}}^2({\rho} -1)^2 +{\rho} ^2), \nonumber \\
		f_{4}(\rho) &= L_\mathrm g{\rho}^2+CK_{\mathrm{b}}^2({\rho} -1)^2(CL_\mathrm g-1), \nonumber \\
		f_{5}(\rho) &= (CL_\mathrm g-1)^2K_{\mathrm{b}}^2({\rho} -1)^2+(CR_\mathrm g)^2K_{\mathrm{b}}^2({\rho} -1)^2 \nonumber\\&-2K_{\mathrm{b}}R_\mathrm g{\rho}({\rho} -1) +{\rho} ^2(R_\mathrm g^2+L_\mathrm g^2). \nonumber
	\end{align*}
	
	\subsubsection{Inductive Interconnecting Lines}
	\label{eq:reduced-model-L}
	The dynamics of the slow-varying states are given by
	\begin{subequations}
		\begin{align}
			\dot{\theta} &= \omega_\mathrm b + \frac{\omega_{\mathrm{b}}{\kappa_1}}{(E^{\star})^2}\mathrm{e}_{1}^\top \mathrm{T}_{2}(\psi - \tfrac{\pi}{2})(S^\star - S), \label{eq:differential4thorder_inductive1} \\
			\dot{E}^{\star} &= \frac{\omega_{\mathrm{b}}{\kappa_1}}{{E^{\star}}}\mathrm{e}_{2}^\top \mathrm{T}_{2}(\psi - \tfrac{\pi}{2})(S^\star - S) \nonumber\\ &+ \omega_{\mathrm{b}}\kappa_2(E_{\mathrm{b}}^2-(E^{\star})^2)E^{\star}, \label{eq:differential4thorder_inductive2}\\
			\dot{{I}}_{\mathrm{g}dq} &= \omega_\mathrm b \Big(\mathrm{T}_{2}(\tfrac{\pi}{2})\Big(\mathrm I-\frac{1}{L_\mathrm gC}\left(\mathrm I-\rho \mathrm{A}_{1}(\rho)\right)\Big)  - \frac{R_\mathrm g}{L_\mathrm g}\mathrm I\Big) I_{\mathrm{g}dq} \nonumber \\
			&+ \frac{\omega_\mathrm b}{L_\mathrm g} \left( \frac{\rho}{C}\mathrm{T}_{2}(\tfrac{\pi}{2})\mathrm{A}_{2}(\rho)\mathrm{e}_1{E}^{\star} - \mathrm{T}_{2}(\delta)V_{DQ}\right),  \label{eq:differential4thorder_inductive3}
		\end{align}
	\end{subequations}
	where $\rho$ is given by the solution of \eqref{eq:rho4thorder} and the active- and reactive-power values, $P,Q$ in $S = [P,Q]^\top$ take the form:
	\begin{subequations}
		\begin{align}
			P=&\ I_{\mathrm{g}dq}^{\top}\Big(\frac{\rho}{C} \mathrm{A}_{1}(\rho)^{\top}\mathrm{T}_{2}(\tfrac{\pi}{2})^{\top}
			-\frac{1}{C}\mathrm{T}_{2}(\tfrac{\pi}{2})^{\top}\Big)I_{\mathrm{g}dq}  \nonumber\\
			&+\frac{\rho}{C} \mathrm{e}_1^{\top}\mathrm{A}_{2}(\rho)^{\top}\mathrm{T}_{2}(\tfrac{\pi}{2})^{\top} E^{\star}I_{\mathrm{g}dq},\label{eq:P}\\
			Q=&\ I_{\mathrm{g}dq}^{\top}\Big(\frac{1}{C}\mathrm I -\frac{\rho}{C} \mathrm{A}_{1}(\rho)^{\top}\Big)I_{\mathrm{g}dq}\nonumber\\ &-\frac{\rho}{C} \mathrm{e}_1^{\top}\mathrm{A}_{2}(\rho)^{\top} E^{\star}I_{\mathrm{g}dq}.\label{eq:Q}
		\end{align}
	\end{subequations}
	Algebraic equations for the fast-varying state variables are:
	\begin{subequations}\label{eq:manifold-1}
		\begin{align}
			{{I}_{{\mathrm{i}}dq}}&= {\rho} (\mathrm{A}_{1}(\rho)I_{\mathrm{g}dq} + \mathrm{A}_{2}(\rho)\mathrm{e}_1E^{\star}), \label{eqn:4th-order5a}\\
			{E}_{dq}&= \frac{1}{C} \mathrm{T}_{2}(\tfrac{\pi}{2}) \Big({{I}_{{\mathrm{i}}dq}}-{I}_{{\mathrm{g}}dq}\Big),\label{eqn:4th-order5b}\\
			{\Phi}_{dq}&= \frac{1}{{\rho}K_{\mathrm{I}v}}({\rho} -1)(K_{\mathrm{b}}K_{\mathrm{P}v}-1){I}_{{\mathrm{i}}dq},  \label{eqn:4th-order5c}\\ 
			{\Gamma}_{dq}&=  \frac{{R}_{\mathrm{i}} }{K_{\mathrm{I}i}}{{I}_{{\mathrm{i}}dq}}. \label{eqn:4th-order5d}
		\end{align} 
	\end{subequations} Note that the equations comprising \eqref{eq:manifold-1} can be expressed as functions of $\delta$, $E^{\star}$, ${I}_{{\mathrm{g}}dq}$, and $\rho$.
	
	\subsubsection{Resistive Interconnecting Lines}
	\label{eq:reduced-model-R}
	The dynamics of $\theta$, $E^\star$ are the same as in~\eqref{eq:differential4thorder_inductive1}--\eqref{eq:differential4thorder_inductive2}; $P,Q$ in $S=[P,\,\,Q]^\top$ remain defined as in~\eqref{eq:P}--\eqref{eq:Q}; and $\rho$ is still defined by the solution of \eqref{eq:rho4thorder}. However, $I_{\mathrm{g}dq}$, which is now a fast-varying state, is defined algebraically via
	\begin{align} \label{eq:IGDQ_algebraic}
		{I}_{{\mathrm{g}}dq}&= \mathrm{A}_{3}(\rho)\mathrm{e}_1{E}^{\star} + \mathrm{A}_{4}(\rho)\mathrm{T}_{2}(\delta)V_{DQ}.
	\end{align} 
	The algebraic equations for the other fast-varying state variables are the same as~\eqref{eqn:4th-order5a}--\eqref{eqn:4th-order5d}. 
	
	\begin{remark}[Structure of Reduced-order Models]
		The reduced-order models for the slow-varying states in both cases considered above are self contained, in that they do not invoke any fast-varying states. The dynamics are DAE models in each case, with the algebraic component given by~\eqref{eq:rho4thorder}. 
	\end{remark}
	
	\subsection{Special Case with Current-reference Limiter Ignored} \label{sec:rhoequalone}
	The current-reference limiter can be ignored in the dynamical model presented in Section~\ref{subsec:ModRed} by setting~$\rho = 1$. The collection of slow and fast states for the inductive and resistive interconnecting lines remains the same as before.
	
	For inductive interconnecting lines, the dynamics of $\theta$ and $E^\star$ are the same as~\eqref{eq:delta}--\eqref{eq:Estar}, except, with the active- and reactive-power values, $P,Q$ simplifying to:
	$P= \mathrm{e}_1^{\top} E^{\star}I_{\mathrm{g}dq}$, 
	$Q=-\mathrm{e}_2^{\top} E^{\star}I_{\mathrm{g}dq}$.
	The dynamics of $I_{\mathrm{g}dq}$ are given by{
		\begin{align}
			\begin{split}
				\dot{{I}}_{\mathrm{g}dq} &= \omega_\mathrm b\left(\mathrm{T}_{2}(\tfrac{\pi}{2}) - \frac{R_\mathrm g}{L_\mathrm g}\mathrm I\right) I_{\mathrm{g}dq} + \frac{\omega_\mathrm b}{L_\mathrm g} \left( \mathrm{e}_1{E}^{\star} - \mathrm{T}_{2}(\delta)V_{DQ}\right).\nonumber
			\end{split}
	\end{align}}
	The algebraic equations for the fast-varying state variables are:
	\begin{equation*}
		\begin{split}
			{{I}_{{\mathrm{i}}dq}}&= C\mathrm{e}_2E^{\star} + I_{\mathrm{g}dq}, \qquad
			{E}_{dq}= \mathrm{e}_1E^{\star},\\
			{\Phi}_{dq}&= [0\,, 0]^\top,\qquad
			{\Gamma}_{dq}=  \frac{{R}_{\mathrm{i}} }{K_{\mathrm{I}i}} (C\mathrm{e}_2E^{\star} + I_{\mathrm{g}dq}).
		\end{split}
	\end{equation*}
	From above, we see that the filter-capacitor voltage, $E_{dq}$, is regulated to the reference generated by the AHO model, $\mathrm{e}_1 E^\star$. 
	
	For resistive interconnecting lines, the dynamics of $\theta$, $E^\star$ are the same as reported above for inductive interconnecting lines, as are algebraic constraints for $\,{I}_{\mathrm{i}dq},\,{E}_{dq},\,\Phi_{dq},\,\Gamma_{dq}$. The algebraic constraint for $I_{\mathrm{g}dq}$ can be recovered from~\eqref{eq:IGDQ_algebraic} by substituting $\mathrm{A}_{3}(1)$ and $\mathrm{A}_{4}(1)$.
	
	\section{Numerical Results}\label{sec:simRes}
\begin{figure}[t!]
		\centering
		\includegraphics{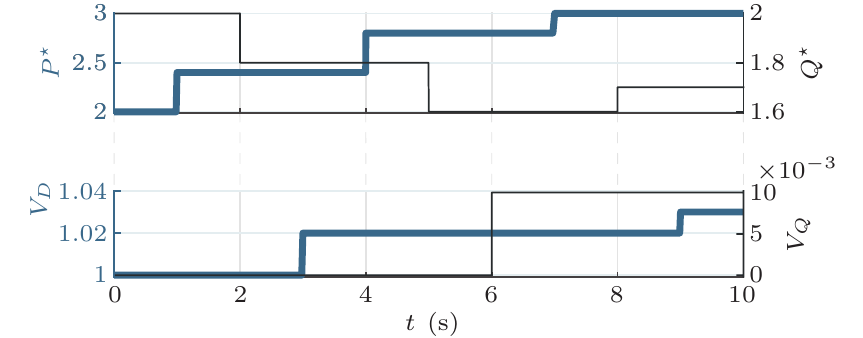}
		\caption{Profile of power-reference inputs and $DQ$ components of the grid-side voltage utilized in the simulations.}
		\label{fig:references}
		\vspace{-0.2in}
	\end{figure}

	We consider a GFM inverter with dVOC interconnected to an infinite bus via: i) an inductive line, and ii) a resistive line. The active- and reactive-power references, and infinite-bus voltage are varied over a $10$~\si{\second} time interval according to Fig.~\ref{fig:references}. We include simulations from a switched version of the model in Fig.~\ref{fig:AHOinverter}, the averaged full-order model discussed in Section~\ref{sec:full-order}, and the reduced-order models discussed in Section~\ref{subsec:ModRed}. Parameters in Table~\ref{tab:param} are used in all simulations.  
	
\begin{figure*}[t]
		\centering
		\includegraphics{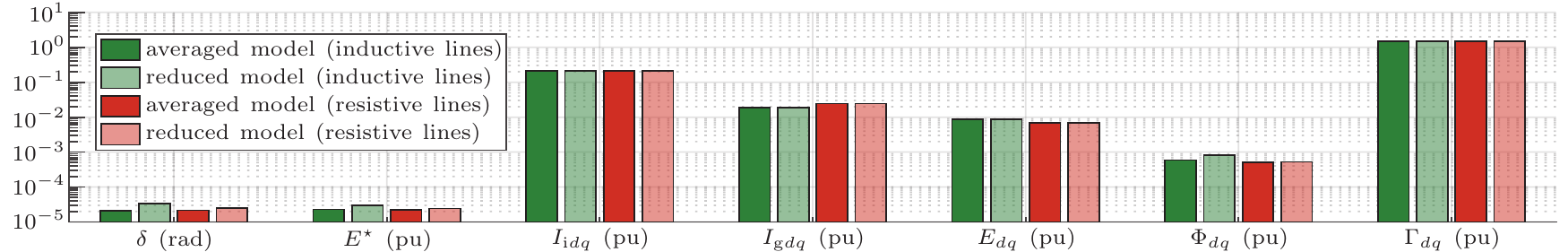}
		\caption{RMSE of the averaged full-order model and the reduced-order model relative to the switched model.}
		\label{fig:rmse_log}
	\end{figure*}

\begin{figure*}
		\centering
		\begin{subfigure}{\columnwidth}
			\centering
			\includegraphics{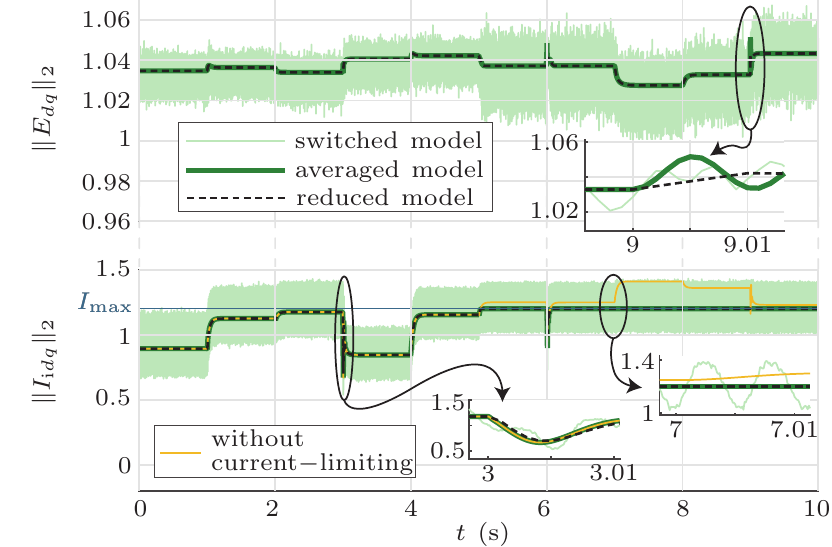}
			\caption{With inductive interconnecting lines.}
			\label{fig:TimeScalePlots_L}
		\end{subfigure}
		\hfill
		\begin{subfigure}{\columnwidth}
			\centering
			\includegraphics{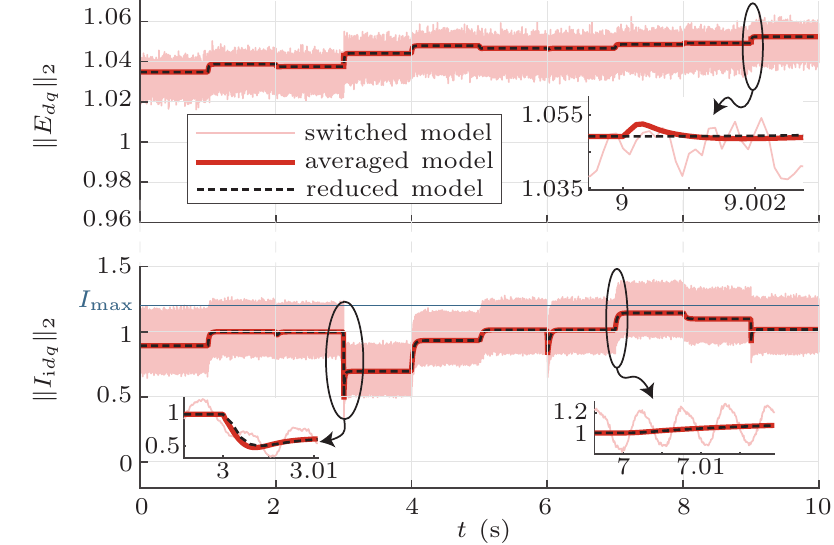}
			\caption{With resistive interconnecting lines.}
			\label{fig:TimeScalePlots_R}
		\end{subfigure}
		\caption{{Output response of the GFM inverter's switched model, averaged full-order model, and reduced-order model.}} \vspace{-0.2in}
	\end{figure*}

	The computational effort required by the switched full-order model, the averaged full-order model, and the reduced-order model are 40.5~\si{\second}, 14.3~\si{\second}, and 1.9~\si{\second}, respectively. Figure~\ref{fig:rmse_log} shows the root mean square error (RMSE) of the averaged full-order model's response and the reduced-order model's response, relative to the switched full-order model. Additionally, Figs.~\ref{fig:TimeScalePlots_L} and~\ref{fig:TimeScalePlots_R} depict the output voltage and current of the switched model, the averaged full-order model, and the reduced model. Note that the reduced-order models capture the effects of the current-reference limiter. Understandably, the reduced-order models do not capture all higher-order transients, but they do preserve all dominant transient behavior and return the same steady-state values as the higher-order models. The numerical results show that, although the RMSE associated with the reduced-order models and the full-order model have identical orders of magnitude, our proposed reduced-order models require an order-of-magnitude less computational effort.
	
	
	\section{Concluding Remarks \& Future Work}\label{sec:conclusion}
	This work outlined reduced-order models for grid-forming inverters realized with dispatchable Virtual Oscillator Control. Compared to previous efforts for model reduction, our proposed models  retain the effects of the current-reference limiter in the model response. Simulation results indicate that the proposed reduced-order models require an order-of-magnitude less computational effort to produce results with the same order-of-magnitude accuracy as the averaged full-order model. Future work includes extending the results presented in this paper to other types of GFM inverters.
	
	\vspace{-0.03in}
	\bibliographystyle{IEEEtran}
	\bibliography{references}
	
	\begin{appendix}[{Model-order Reduction Steps}]
		
		\subsection{The 12th-order Model} Following step~1 of Section~\ref{subsec:ModRed}, the dynamics introduced in Section~\ref{sec:full-order} can be expressed compactly as follows:
		\begin{subequations}\label{eq:12th-order}
			\begin{align}	
				\dot{\delta} &= \frac{\omega_{\mathrm{b}} \kappa_1}{(E^{\star})^2} \mathrm{e}_{1}^\top \mathrm{T}_{2}(\psi - \tfrac{\pi}{2})(S^\star - S), \label{eq:delta-X}\\
				\dot{E}^{\star}\! &= \frac{\omega_{\mathrm{b}}{\kappa_1}}{E^{\star}}  \mathrm{e}_{2}^\top \mathrm{T}_{2}(\psi - \tfrac{\pi}{2})(S^\star - S) \nonumber \\ & \qquad\qquad + \omega_{\mathrm{b}}\kappa_2(E_{\mathrm{b}}^2-(E^{\star})^2)E^{\star}, \label{eq:Estar-X}\\
				\dot{{I}}_{\mathrm{g}dq} &= \Big(\omega \mathrm{T}_{2}(\tfrac{\pi}{2}) - \omega_\mathrm b\frac{R_\mathrm g}{L_\mathrm g}\mathrm I\Big) I_{\mathrm{g}dq} + \frac{\omega_\mathrm b}{L_\mathrm g} ( E_{dq} - \mathrm{T}_{2}(\delta)V_{DQ}), \label{eq:LCL_filter3-X}\\
				\dot{{I}}_{\mathrm{i}dq} &= - \frac{\omega_\mathrm b(R_\mathrm i+K_{\mathrm{P}i})}{L_\mathrm i}I_{\mathrm{i}dq} + \frac{\omega_\mathrm b}{L_\mathrm i} \Big(K_{\mathrm{P}i}\Big(\rho\Big(K_{\mathrm{P}v}\mathrm{e}_{1}{E}^{\star}\nonumber \\ & + {I}_{\mathrm{g}dq} + K_{\mathrm{I}v}
				{\Phi}_{dq} - \Big(K_{\mathrm{P}v}\mathrm I + \frac{\omega}{\omega_\mathrm b}C \mathrm{T}_{2}(\tfrac{\pi}{2})\Big)E_{dq}\Big)\Big)\nonumber \\ & + K_{\mathrm{I}i}
				{\Gamma}_{dq}\Big),\label{eq:LCL_filter1-X} \\
				\dot{E}_{dq} &= \omega \mathrm{T}_{2}(\tfrac{\pi}{2}) E_{dq} + \frac{\omega_\mathrm b}{C} (I_{\mathrm{i}dq} - I_{\mathrm{g}dq}), \label{eq:LCL_filter2-X}\\
				\dot{\Phi}_{dq}
				&= \omega_{\mathrm{b}}(1+K_{\mathrm{b}}(\rho-1)K_{\mathrm{P}v})\mathrm{e}_{1}{E}^{\star} \nonumber \\ &-\omega_{\mathrm{b}}\Big(\mathrm I + K_{\mathrm{b}}(\rho-1)\Big(K_{\mathrm{P}v}\mathrm I + \frac{\omega}{\omega_\mathrm b}C \mathrm{T}_{2}(\tfrac{\pi}{2})\Big)\Big)E_{dq}\nonumber \\ &+ \omega_{\mathrm{b}}K_{\mathrm{b}}(\rho-1)( K_{\mathrm{I}v}
				{\Phi}_{dq} + {I}_{\mathrm{g}dq}),\label{eq:Vcont1-X}\\
				\dot{\Gamma}_{dq} &= \omega_\mathrm b \rho \Big(K_{\mathrm{P}v}\mathrm{e}_{1}{E}^{\star} + K_{\mathrm{I}v}
				{\Phi}_{dq} + {I}_{\mathrm{g}dq} \nonumber\\& - \Big(K_{\mathrm{P}v}\mathrm I + \frac{\omega}{\omega_\mathrm b}C \mathrm{T}_{2}(\tfrac{\pi}{2})\Big)E_{dq}\Big) - \omega_\mathrm b I_{\mathrm{i}dq}, \label{eq:Gamma-X}\\
				\rho &=-\varepsilon\ln\Bigg(\exp\left(\frac{-1}{\varepsilon}\right)+\exp\bigg({-I_\mathrm{max}} \div\varepsilon\Big\lVert K_{\mathrm{P}v}\mathrm{e}_{1}{E}^{\star} \nonumber\\ & + K_{\mathrm{I}v}
				{\Phi}_{dq} + {I}_{\mathrm{g}dq} - \Big(K_{\mathrm{P}v}\mathrm I + \frac{\omega}{\omega_\mathrm b}C \mathrm{T}_{2}(\tfrac{\pi}{2})\Big)E_{dq}\Big\rVert_2\bigg)\Bigg), \label{eq:rho-X}\\
				\omega &= \omega_{\mathrm{b}} + \frac{\omega_{\mathrm{b}} \kappa_1}{(E^{\star})^2} \mathrm{e}_{1}^\top \mathrm{T}_{2}(\psi - \tfrac{\pi}{2})(S^\star - S),\label{eq:omega-X}
			\end{align}
		\end{subequations} where $S =[P, Q]^\top$, with $P$ and $Q$ described by \eqref{eq:PQ}.
		
		\subsection{The Standard Singular Perturbation Form}
		Following step~2 of Section~\ref{subsec:ModRed}, with inductive lines interconnecting the GFM inverter to the bus, the dynamics in \eqref{eq:12th-order} can be expressed in the standard singular perturbation form as follows:
		\begin{subequations}\label{eq:standard-form-L}
			\begin{align}	
				\dot{\delta} &= \frac{\omega_{\mathrm{b}} \kappa_1}{(E^{\star})^2} \mathrm{e}_{1}^\top \mathrm{T}_{2}(\psi - \tfrac{\pi}{2})(S^\star - S), \label{eq:delta-L}\\
				\dot{E}^{\star}\! &= \frac{\omega_{\mathrm{b}}{\kappa_1}}{E^{\star}}  \mathrm{e}_{2}^\top \mathrm{T}_{2}(\psi - \tfrac{\pi}{2})(S^\star - S) \nonumber \\ & \qquad\qquad + \omega_{\mathrm{b}}\kappa_2(E_{\mathrm{b}}^2-(E^{\star})^2)E^{\star}, \label{eq:Estar-L}\\
				\dot{{I}}_{\mathrm{g}dq} &= \Big(\omega \mathrm{T}_{2}(\tfrac{\pi}{2}) - \omega_\mathrm b\frac{R_\mathrm g}{L_\mathrm g}\mathrm I\Big) I_{\mathrm{g}dq} + \frac{\omega_\mathrm b}{L_\mathrm g} ( E_{dq} - \mathrm{T}_{2}(\delta)V_{DQ}), \label{eq:LCL_filter3-L}\\
				\epsilon\dot{{I}}_{\mathrm{i}dq} &= - \epsilon\frac{\omega_\mathrm b(R_\mathrm i+K_{\mathrm{P}i})}{L_\mathrm i}I_{\mathrm{i}dq} + \epsilon\frac{\omega_\mathrm b}{L_\mathrm i} \Big(K_{\mathrm{P}i}\Big(\rho\Big(K_{\mathrm{P}v}\mathrm{e}_{1}{E}^{\star}\nonumber \\ & + {I}_{\mathrm{g}dq} + K_{\mathrm{I}v}
				{\Phi}_{dq} - \Big(K_{\mathrm{P}v}\mathrm I + \frac{\omega}{\omega_\mathrm b}C \mathrm{T}_{2}(\tfrac{\pi}{2})\Big)E_{dq}\Big)\Big)\nonumber \\ & + K_{\mathrm{I}i}
				{\Gamma}_{dq}\Big),\label{eq:LCL_filter1-L} \\
				\epsilon\dot{E}_{dq} &= \epsilon\omega \mathrm{T}_{2}(\tfrac{\pi}{2}) E_{dq} + \epsilon\frac{\omega_\mathrm b}{C} (I_{\mathrm{i}dq} - I_{\mathrm{g}dq}), \label{eq:LCL_filter2-L}\\
				\epsilon\dot{\Phi}_{dq}
				&= \epsilon\omega_{\mathrm{b}}(1+K_{\mathrm{b}}(\rho-1)K_{\mathrm{P}v})\mathrm{e}_{1}{E}^{\star} \nonumber \\ &-\epsilon\omega_{\mathrm{b}}\Big(\mathrm I + K_{\mathrm{b}}(\rho-1)\Big(K_{\mathrm{P}v}\mathrm I + \frac{\omega}{\omega_\mathrm b}C \mathrm{T}_{2}(\tfrac{\pi}{2})\Big)\Big)E_{dq}\nonumber \\ &+ \epsilon\omega_{\mathrm{b}}K_{\mathrm{b}}(\rho-1)( K_{\mathrm{I}v}
				{\Phi}_{dq} + {I}_{\mathrm{g}dq}),\label{eq:Vcont1-L}\\
				\epsilon\dot{\Gamma}_{dq} &= \epsilon\omega_\mathrm b \rho \Big(K_{\mathrm{P}v}\mathrm{e}_{1}{E}^{\star} + K_{\mathrm{I}v}
				{\Phi}_{dq} + {I}_{\mathrm{g}dq} \nonumber\\& - \Big(K_{\mathrm{P}v}\mathrm I + \frac{\omega}{\omega_\mathrm b}C \mathrm{T}_{2}(\tfrac{\pi}{2})\Big)E_{dq}\Big) - \epsilon\omega_\mathrm b I_{\mathrm{i}dq}, \label{eq:Gamma-L}\\
				\rho &=-\varepsilon\ln\Bigg(\exp\left(\frac{-1}{\varepsilon}\right)+\exp\bigg({-I_\mathrm{max}} \div\varepsilon\Big\lVert K_{\mathrm{P}v}\mathrm{e}_{1}{E}^{\star} \nonumber\\ & + K_{\mathrm{I}v}
				{\Phi}_{dq} + {I}_{\mathrm{g}dq} - \Big(K_{\mathrm{P}v}\mathrm I + \frac{\omega}{\omega_\mathrm b}C \mathrm{T}_{2}(\tfrac{\pi}{2})\Big)E_{dq}\Big\rVert_2\bigg)\Bigg), \label{eq:rho-L}\\
				\omega &= \omega_{\mathrm{b}} + \frac{\omega_{\mathrm{b}} \kappa_1}{(E^{\star})^2} \mathrm{e}_{1}^\top \mathrm{T}_{2}(\psi - \tfrac{\pi}{2})(S^\star - S),\label{eq:omega-L}
			\end{align}
		\end{subequations} and when resistive lines interconnect the GFM inverter to the bus, the dynamics in \eqref{eq:12th-order} can be expressed in the standard singular perturbation form as follows:
		\begin{subequations}\label{eq:standard-form-R}
			\begin{align}	
				\dot{\delta} &= \frac{\omega_{\mathrm{b}} \kappa_1}{(E^{\star})^2} \mathrm{e}_{1}^\top \mathrm{T}_{2}(\psi - \tfrac{\pi}{2})(S^\star - S), \label{eq:delta-R}\\
				\dot{E}^{\star}\! &= \frac{\omega_{\mathrm{b}}{\kappa_1}}{E^{\star}}  \mathrm{e}_{2}^\top \mathrm{T}_{2}(\psi - \tfrac{\pi}{2})(S^\star - S) \nonumber \\ & \qquad\qquad + \omega_{\mathrm{b}}\kappa_2(E_{\mathrm{b}}^2-(E^{\star})^2)E^{\star}, \label{eq:Estar-R}\\
				\epsilon\dot{{I}}_{\mathrm{g}dq} &= \epsilon\Big(\omega \mathrm{T}_{2}(\tfrac{\pi}{2}) - \omega_\mathrm b\frac{R_\mathrm g}{L_\mathrm g}\mathrm I\Big) I_{\mathrm{g}dq}\nonumber \\ & \qquad\qquad  + \epsilon\frac{\omega_\mathrm b}{L_\mathrm g} ( E_{dq} - \mathrm{T}_{2}(\delta)V_{DQ}), \label{eq:LCL_filter3-R}\\
				\epsilon\dot{{I}}_{\mathrm{i}dq} &= - \epsilon\frac{\omega_\mathrm b(R_\mathrm i+K_{\mathrm{P}i})}{L_\mathrm i}I_{\mathrm{i}dq} + \epsilon\frac{\omega_\mathrm b}{L_\mathrm i} \Big(K_{\mathrm{P}i}\Big(\rho\Big(K_{\mathrm{P}v}\mathrm{e}_{1}{E}^{\star}\nonumber \\ & + {I}_{\mathrm{g}dq} + K_{\mathrm{I}v}
				{\Phi}_{dq} - \Big(K_{\mathrm{P}v}\mathrm I + \frac{\omega}{\omega_\mathrm b}C \mathrm{T}_{2}(\tfrac{\pi}{2})\Big)E_{dq}\Big)\Big)\nonumber \\ & + K_{\mathrm{I}i}
				{\Gamma}_{dq}\Big),\label{eq:LCL_filter1-R} \\
				\epsilon\dot{E}_{dq} &= \epsilon\omega \mathrm{T}_{2}(\tfrac{\pi}{2}) E_{dq} + \epsilon\frac{\omega_\mathrm b}{C} (I_{\mathrm{i}dq} - I_{\mathrm{g}dq}), \label{eq:LCL_filter2-R}\\
				\epsilon\dot{\Phi}_{dq}
				&= \epsilon\omega_{\mathrm{b}}(1+K_{\mathrm{b}}(\rho-1)K_{\mathrm{P}v})\mathrm{e}_{1}{E}^{\star} \nonumber \\ &-\epsilon\omega_{\mathrm{b}}\Big(\mathrm I + K_{\mathrm{b}}(\rho-1)\Big(K_{\mathrm{P}v}\mathrm I + \frac{\omega}{\omega_\mathrm b}C \mathrm{T}_{2}(\tfrac{\pi}{2})\Big)\Big)E_{dq}\nonumber \\ &+ \epsilon\omega_{\mathrm{b}}K_{\mathrm{b}}(\rho-1)( K_{\mathrm{I}v}
				{\Phi}_{dq} + {I}_{\mathrm{g}dq}),\label{eq:Vcont1-R}\\
				\epsilon\dot{\Gamma}_{dq} &= \epsilon\omega_\mathrm b \rho \Big(K_{\mathrm{P}v}\mathrm{e}_{1}{E}^{\star} + K_{\mathrm{I}v}
				{\Phi}_{dq} + {I}_{\mathrm{g}dq} \nonumber\\& - \Big(K_{\mathrm{P}v}\mathrm I + \frac{\omega}{\omega_\mathrm b}C \mathrm{T}_{2}(\tfrac{\pi}{2})\Big)E_{dq}\Big) - \epsilon\omega_\mathrm b I_{\mathrm{i}dq}, \label{eq:Gamma-R}\\
				\rho &=-\varepsilon\ln\Bigg(\exp\left(\frac{-1}{\varepsilon}\right)+\exp\bigg({-I_\mathrm{max}} \div\varepsilon\Big\lVert K_{\mathrm{P}v}\mathrm{e}_{1}{E}^{\star} \nonumber\\ & + K_{\mathrm{I}v}
				{\Phi}_{dq} + {I}_{\mathrm{g}dq} - \Big(K_{\mathrm{P}v}\mathrm I + \frac{\omega}{\omega_\mathrm b}C \mathrm{T}_{2}(\tfrac{\pi}{2})\Big)E_{dq}\Big\rVert_2\bigg)\Bigg), \label{eq:rho-R}\\
				\omega &= \omega_{\mathrm{b}} + \frac{\omega_{\mathrm{b}} \kappa_1}{(E^{\star})^2} \mathrm{e}_{1}^\top \mathrm{T}_{2}(\psi - \tfrac{\pi}{2})(S^\star - S),\label{eq:omega-R}
			\end{align}
		\end{subequations}

		\subsection{The Reduced-order Models}
		Following step~3 of Section~\ref{subsec:ModRed}, when inductive interconnecting lines are used, the reduced-order model is derived by:
		\begin{enumerate}[(i)]
			\item Setting $\epsilon=0$ on the left-hand sides of \eqref{eq:LCL_filter1-L}--\eqref{eq:Gamma-L} and setting $\omega = \omega_{\mathrm{b}}$ in \eqref{eq:omega-L} to get
			\begin{subequations}\label{eq:standard-form-L1}
				\begin{align}	
					0 &= - \frac{\omega_\mathrm b(R_\mathrm i+K_{\mathrm{P}i})}{L_\mathrm i}I_{\mathrm{i}dq} + \frac{\omega_\mathrm b}{L_\mathrm i} \Big(K_{\mathrm{P}i}\Big(\rho\Big(K_{\mathrm{P}v}\mathrm{e}_{1}{E}^{\star}\nonumber \\ & + K_{\mathrm{I}v}
					{\Phi}_{dq} + {I}_{\mathrm{g}dq}  - \Big(K_{\mathrm{P}v}\mathrm I + C \mathrm{T}_{2}(\tfrac{\pi}{2})\Big)E_{dq}\Big)\Big)\nonumber \\ & + K_{\mathrm{I}i}
					{\Gamma}_{dq}\Big),\label{eq:LCL_filter1-L1} \\
					0 &= \omega_{\mathrm{b}}\mathrm{T}_{2}(\tfrac{\pi}{2}) E_{dq} + \frac{\omega_\mathrm b}{C} (I_{\mathrm{i}dq} - I_{\mathrm{g}dq}), \label{eq:LCL_filter2-L1}\\
					0
					&= \omega_{\mathrm{b}}(1+K_{\mathrm{b}}(\rho-1)K_{\mathrm{P}v})\mathrm{e}_{1}{E}^{\star} \nonumber \\ &-\omega_{\mathrm{b}}\Big(\mathrm I + K_{\mathrm{b}}(\rho-1)\Big(K_{\mathrm{P}v}\mathrm I + C \mathrm{T}_{2}(\tfrac{\pi}{2})\Big)\Big)E_{dq}\nonumber \\ &+ \omega_{\mathrm{b}}K_{\mathrm{b}}(\rho-1)( K_{\mathrm{I}v}
					{\Phi}_{dq} + {I}_{\mathrm{g}dq}),\label{eq:Vcont1-L1}\\
					0 &= \omega_\mathrm b \rho \Big(K_{\mathrm{P}v}\mathrm{e}_{1}{E}^{\star} + K_{\mathrm{I}v}
					{\Phi}_{dq} + {I}_{\mathrm{g}dq} \nonumber\\& - \Big(K_{\mathrm{P}v}\mathrm I + C \mathrm{T}_{2}(\tfrac{\pi}{2})\Big)E_{dq}\Big) - \omega_\mathrm b I_{\mathrm{i}dq}, \label{eq:Gamma-L1}\\
					\rho &=-\varepsilon\ln\Bigg(\exp\left(\frac{-1}{\varepsilon}\right)+\exp\bigg({-I_\mathrm{max}} \div\varepsilon\Big\lVert K_{\mathrm{P}v}\mathrm{e}_{1}{E}^{\star} \nonumber\\ & + K_{\mathrm{I}v}
					{\Phi}_{dq} + {I}_{\mathrm{g}dq} - \Big(K_{\mathrm{P}v}\mathrm I + C \mathrm{T}_{2}(\tfrac{\pi}{2})\Big)E_{dq}\Big\rVert_2\bigg)\Bigg). \label{eq:rho-L1}
				\end{align}
			\end{subequations}
			\item Noting that $\mathrm{T}_{2}^{-1}(\tfrac{\pi}{2})=-\mathrm{T}_{2}(\tfrac{\pi}{2})$, from \eqref{eq:LCL_filter2-L1} and solving for ${E}_{dq}$ to get
			\begin{align}
				E_{dq} =  \frac{1}{C} \mathrm{T}_{2}(\tfrac{\pi}{2})(I_{\mathrm{i}dq} - I_{\mathrm{g}dq}),\label{eq:LCL_filter2-L1X}
			\end{align} which is the expression presented in \eqref{eqn:4th-order5b}.
			\item Noting that from \eqref{eq:Gamma-L1} we have that
			\begin{align}
				\frac{I_{\mathrm{i}dq}}{\rho} &= K_{\mathrm{P}v}\mathrm{e}_{1}{E}^{\star} + K_{\mathrm{I}v}
				{\Phi}_{dq} + {I}_{\mathrm{g}dq} - \big(K_{\mathrm{P}v}\mathrm I \nonumber \\&+ C \mathrm{T}_{2}(\tfrac{\pi}{2})\big)E_{dq}, \label{eq:Gamma-L1X}
			\end{align} and from \eqref{eq:Vcont1-L1} we have that
			\begin{align}
				0
				&= \mathrm{e}_{1}{E}^{\star}-E_{dq}+K_{\mathrm{b}}(\rho-1)\Big(K_{\mathrm{P}v}\mathrm{e}_{1}{E}^{\star} \nonumber \\ & + K_{\mathrm{I}v}
				{\Phi}_{dq} + {I}_{\mathrm{g}dq} - \big(K_{\mathrm{P}v}\mathrm I + C \mathrm{T}_{2}(\tfrac{\pi}{2})\big)E_{dq} \Big), \label{eq:Vcont1-L1X}
			\end{align} substituting \eqref{eq:Gamma-L1X} into \eqref{eq:Vcont1-L1X} to get
			\begin{align}
				0
				&= \mathrm{e}_{1}{E}^{\star}-E_{dq}+K_{\mathrm{b}}(\rho-1)\frac{I_{\mathrm{i}dq}}{\rho},\label{eq:Vcont1-L1XX}
			\end{align} substituting \eqref{eq:LCL_filter2-L1X} into \eqref{eq:Vcont1-L1XX} to get
			\begin{align}
				\Big(\frac{{\rho}}{C} \mathrm{T}_{2}(\tfrac{\pi}{2}) &- {K_{\mathrm{b}}(\rho-1)}\mathrm{I}\Big){I_{\mathrm{i}dq}}
				\nonumber\\ &= {\rho}\Big(\frac{1}{C} \mathrm{T}_{2}(\tfrac{\pi}{2})I_{\mathrm{g}dq} + \mathrm{e}_{1}{E}^{\star}\Big).\label{eq:Vcont1-L1XXX}
			\end{align} Noting that
			\begin{align*}
				\mathrm{A}_{1}(\rho) &= \Big(\frac{{\rho}}{C} \mathrm{T}_{2}(\tfrac{\pi}{2}) - {K_{\mathrm{b}}(\rho-1)}\mathrm{I}\Big)^{-1}\frac{1}{C} \mathrm{T}_{2}(\tfrac{\pi}{2}),\\
				\mathrm{A}_{2}(\rho) &= \Big(\frac{{\rho}}{C} \mathrm{T}_{2}(\tfrac{\pi}{2}) - {K_{\mathrm{b}}(\rho-1)}\mathrm{I}\Big)^{-1},
			\end{align*} and solving for ${I_{\mathrm{i}dq}}$ using \eqref{eq:Vcont1-L1XXX} to get
			\begin{align}
				{{I}_{{\mathrm{i}}dq}} &= {\rho} (\mathrm{A}_{1}(\rho)I_{\mathrm{g}dq} + \mathrm{A}_{2}(\rho)\mathrm{e}_1E^{\star}), \label{eqn:Ii-XX}
			\end{align}
			which is the expression in \eqref{eqn:4th-order5a}.
			\item Substituting \eqref{eq:Gamma-L1X} into \eqref{eq:LCL_filter1-L1} to get
			\begin{align*}	
				0 &= - \frac{\omega_\mathrm b(R_\mathrm i+K_{\mathrm{P}i})}{L_\mathrm i}I_{\mathrm{i}dq} + \frac{\omega_\mathrm b}{L_\mathrm i} \Big(K_{\mathrm{P}i}{I_{\mathrm{i}dq}} + K_{\mathrm{I}i}
				{\Gamma}_{dq}\Big),
			\end{align*} from where it follows that
			\begin{align}	
				{\Gamma}_{dq} &= \frac{R_\mathrm i}{K_{\mathrm{I}i}} I_{\mathrm{i}dq},\label{eq:LCL_filter1-L1X}
			\end{align} which is the expression presented in \eqref{eqn:4th-order5d}.
			\item Substituting \eqref{eq:LCL_filter2-L1X} into \eqref{eq:Gamma-L1} to get
			\begin{align}
				0 &= \rho \Big(K_{\mathrm{P}v}\mathrm{e}_{1}{E}^{\star}-\frac{K_{\mathrm{P}v}}{C} \mathrm{T}_{2}(\tfrac{\pi}{2})(I_{\mathrm{i}dq} - I_{\mathrm{g}dq})\Big) \nonumber\\& + \rho K_{\mathrm{I}v}
				{\Phi}_{dq} + (\rho - 1) I_{\mathrm{i}dq},\nonumber\\
				&= \rho \Big(K_{\mathrm{P}v}\mathrm{e}_{1}{E}^{\star}-{K_{\mathrm{P}v}}{E}_{dq}\Big) + \rho K_{\mathrm{I}v}
				{\Phi}_{dq} \nonumber\\&  + (\rho - 1) I_{\mathrm{i}dq}. \label{eq:Phi-XX}
			\end{align} Noting that from \eqref{eq:Vcont1-L1XX}, $\mathrm{e}_{1}{E}^{\star}-E_{dq}
			= -K_{\mathrm{b}}(\rho-1)\frac{I_{\mathrm{i}dq}}{\rho}$, and substituting this expression into \eqref{eq:Phi-XX} to get
			\begin{align}
				\rho K_{\mathrm{I}v}
				{\Phi}_{dq} &= (\rho-1)K_{\mathrm{P}v}K_{\mathrm{b}}{I_{\mathrm{i}dq}} - (\rho - 1) I_{\mathrm{i}dq},\nonumber\\
				&= (\rho-1)(K_{\mathrm{P}v}K_{\mathrm{b}}-1){I_{\mathrm{i}dq}}, \label{eq:Phi-XXX}
			\end{align} from where it follows that
			\begin{align}
				{\Phi}_{dq} &= \frac{1}{\rho K_{\mathrm{I}v}}(\rho-1)(K_{\mathrm{P}v}K_{\mathrm{b}}-1){I_{\mathrm{i}dq}}, \label{eq:Phi-XXXX}
			\end{align} which is the expression presented in \eqref{eqn:4th-order5c}.
			\item Substituting \eqref{eqn:Ii-XX} into \eqref{eq:LCL_filter2-L1X} to get
			\begin{align}
				E_{dq} =  \frac{1}{C} \mathrm{T}_{2}(\tfrac{\pi}{2})\big(-(\mathrm{I}-{\rho}\mathrm{A}_{1}(\rho))I_{\mathrm{g}dq} + {\rho}\mathrm{A}_{2}(\rho)\mathrm{e}_1E^{\star}\big),\label{eq:LCL_filter2-L1XX}
			\end{align} substituting \eqref{eq:LCL_filter2-L1XX} into \eqref{eq:LCL_filter3-L} to get
			\begin{align}
				\dot{{I}}_{\mathrm{g}dq} &= \omega_\mathrm b \Big(\mathrm{T}_{2}(\tfrac{\pi}{2})\Big(\mathrm I-\frac{1}{L_\mathrm gC}\left(\mathrm I-\rho \mathrm{A}_{1}(\rho)\right)\Big)  \nonumber \\&- \frac{R_\mathrm g}{L_\mathrm g}\mathrm I\Big) I_{\mathrm{g}dq} + \frac{\omega_\mathrm b}{L_\mathrm g} \Big( \frac{\rho}{C}\mathrm{T}_{2}(\tfrac{\pi}{2})\mathrm{A}_{2}(\rho)\mathrm{e}_1{E}^{\star} \nonumber \\&- \mathrm{T}_{2}(\delta)V_{DQ}\Big), \label{eq:LCL_filter-XXX}
			\end{align} which is the expression presented in \eqref{eq:differential4thorder_inductive3}.
			\item Substituting \eqref{eq:Gamma-L1X} into \eqref{eq:rho-L} to get
			\begin{align}	
				\rho =-\varepsilon\ln\Bigg(\exp\Big(\frac{-1}{\varepsilon}\Big)+\exp\Bigg(\frac{-I_\mathrm{max}}{\varepsilon\big\lVert \sfrac{I_{\mathrm{i}dq}}{\rho}\big\rVert_2}\Bigg)\Bigg),\label{eq:rho-XX}
			\end{align} substituting \eqref{eqn:Ii-XX} into \eqref{eq:rho-XX} to get
			\begin{align}
				\rho &=-\varepsilon\ln\Bigg(\exp\left(\frac{-1}{\varepsilon}\right)+\exp\bigg({-I_\mathrm{max}} \div\varepsilon\big\lVert \mathrm{A}_{1}(\rho)I_{\mathrm{g}dq} \nonumber\\ & + \mathrm{A}_{2}(\rho)\mathrm{e}_1E^{\star}\big\rVert_2\bigg)\Bigg),\label{eq:rho-XXX}
			\end{align}
			from where it follows that
			\begin{align}
				{\rho} &= -\varepsilon\ln\Bigg(\exp\Big(-\frac{1}{\varepsilon}\Big) \nonumber \\
				&+\exp\Big(-\frac{I_\mathrm{max}{\sqrt{C^2K_{\mathrm{b}}^2({\rho} -1)^2 +{\rho} ^2}}}{\varepsilon\| C\mathrm{e}_2{E}^{\star}+I_{\mathrm{g}dq}\|_2}\Big)\Bigg),\label{eq:rho-XXXX}
			\end{align} which is equivalent to the expression presented in \eqref{eq:rho4thorder}.
		\end{enumerate}
		When resistive interconnecting lines are used, the reduced-order model is derived by taking the following additional steps:
		\begin{enumerate}[(i)]
			\item Setting $\epsilon=0$ on the left-hand side of \eqref{eq:LCL_filter3-R}, setting $\omega = \omega_{\mathrm{b}}$ on its right-hand side, and substituting \eqref{eq:LCL_filter2-L1XX} into the resulting equation to get
			\begin{align}
				\Big(\frac{R_\mathrm g}{L_\mathrm g}&\mathrm I - \mathrm{T}_{2}(\tfrac{\pi}{2})\Big(\mathrm I-\frac{1}{L_\mathrm gC}\left(\mathrm I-\rho \mathrm{A}_{1}(\rho)\right)\Big)\Big) I_{\mathrm{g}dq} = \nonumber \\& + \frac{\rho}{L_\mathrm g C}\mathrm{T}_{2}(\tfrac{\pi}{2})\mathrm{A}_{2}(\rho)\mathrm{e}_1{E}^{\star} - \frac{1}{L_\mathrm g}\mathrm{T}_{2}(\delta)V_{DQ}. \label{eq:Ig-XXX}
			\end{align}
			\item  Noting that
			\begin{align*}
				\mathrm{A}_{3}(\rho) &= \Big(\frac{R_\mathrm g}{L_\mathrm g}\mathrm I - \mathrm{T}_{2}(\tfrac{\pi}{2})\Big(\mathrm I-\frac{1}{L_\mathrm gC}\big(\mathrm I \nonumber \\& \qquad\qquad - \rho \mathrm{A}_{1}(\rho)\big)\Big)\Big)^{-1}\frac{\rho}{L_\mathrm g C}\mathrm{T}_{2}(\tfrac{\pi}{2})\mathrm{A}_{2}(\rho),\\
				\mathrm{A}_{4}(\rho) &= -\Big(\frac{R_\mathrm g}{L_\mathrm g}\mathrm I - \mathrm{T}_{2}(\tfrac{\pi}{2})\Big(\mathrm I-\frac{1}{L_\mathrm gC}\big(\mathrm I \nonumber \\& \qquad\qquad\qquad\qquad\qquad - \rho \mathrm{A}_{1}(\rho)\big)\Big)\Big)^{-1}\frac{1}{L_\mathrm g},
			\end{align*} and solving for ${I_{\mathrm{g}dq}}$ using \eqref{eq:Ig-XXX} to get
			\begin{align}
				{I}_{{\mathrm{g}}dq}&= \mathrm{A}_{3}(\rho)\mathrm{e}_1{E}^{\star} + \mathrm{A}_{4}(\rho)\mathrm{T}_{2}(\delta)V_{DQ}, \label{eqn:Ig-X}
			\end{align}
			which is the expression in \eqref{eq:IGDQ_algebraic}.
		\end{enumerate}
	\end{appendix}
	
\end{document}